\documentclass[amsmath,amssymb,aps,floatfix,twocolumn,prx,preprintnumbers]{revtex4-1}

\usepackage{graphicx}
\usepackage{upgreek}
\usepackage{amsmath}
\usepackage{amssymb}
\usepackage{chngpage}
\usepackage{tabularx}
\usepackage{array,booktabs}
\newcolumntype{C}{>{\centering\arraybackslash}X}
\usepackage{nicefrac}
\usepackage{color,soul}
\usepackage{xcolor}

\usepackage{enumitem}
 
\begin{document}

\title{Elucidating the $^1$H NMR relaxation mechanism in polydisperse polymers and bitumen using measurements, MD simulations, and models}

\author{Philip M. Singer}
\email{ps41@rice.edu}
\author{Arjun Valiya Parambathu}
\author{Xinglin Wang}
\author{Dilip Asthagiri}
\author{Walter G. Chapman}
\author{George J. Hirasaki}

\affiliation{Department of Chemical and Biomolecular Engineering, Rice University, 6100 Main St., Houston, TX 77005, USA}
\author{Marc Fleury}
\affiliation{IFP Energies Nouvelles, 1 Avenue de Bois-Préau, 92852 Rueil-Malmaison, France}


\begin{abstract}
The mechanism behind the $^1$H NMR frequency dependence of $T_1$ and the viscosity dependence of $T_2$ for polydisperse polymers and bitumen remains elusive. We elucidate the matter through NMR relaxation measurements of polydisperse polymers over an extended range of frequencies ($f_0 = 0.01 \leftrightarrow$ 400 MHz) and viscosities ($\eta = 385 \leftrightarrow 102,000$ cP) using $T_{1}$ and $T_2$ in static fields, $T_{1}$ field-cycling relaxometry, and $T_{1\rho}$ in the rotating frame. We account for the anomalous behavior of the log-mean relaxation times $T_{1LM} \propto f_0$ and $T_{2LM} \propto (\eta/T)^{-1/2}$ with a phenomenological model of $^1$H-$^1$H dipole-dipole relaxation which includes a distribution in molecular correlation times and internal motions of the non-rigid polymer branches. We show that the model also accounts for the anomalous $T_{1LM}$ and $T_{2LM}$ in previously reported bitumen measurements. We find that molecular dynamics (MD) simulations of the $T_{1} \propto f_0$ dispersion and $T_2$ of similar polymers simulated over a range of viscosities ($\eta = 1 \leftrightarrow 1,000$ cP) are in good agreement with measurements and the model. The $T_{1} \propto f_0$ dispersion at high viscosities agrees with previously reported MD simulations of heptane confined in a polymer matrix, which suggests a common NMR relaxation mechanism between viscous polydisperse fluids and fluids under confinement, without the need to invoke paramagnetism.
\end{abstract}

\maketitle

\begin{figure}[!ht]
	\begin{center}
\includegraphics[width=1\columnwidth]{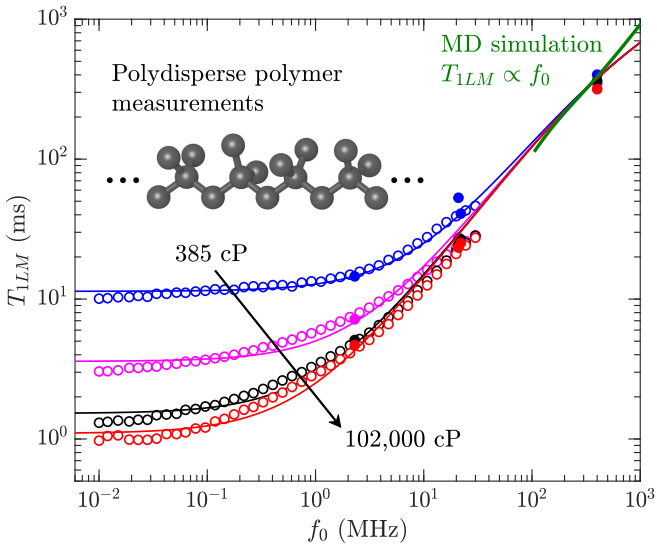}
\end{center}
	\caption{Table of contents; graphical abstract.} \label{fg:Graphical}
\end{figure}

\section{Introduction}\label{sc:Introduction}
Among its many attributes, $^1$H nuclear magnetic resonance (NMR) relaxation is a versatile non-destructive technique for measuring crude-oil viscosity and composition, thus providing a unique contribution to the characterization of light crude-oils, heavy crude-oils, and bitumen \cite{zega:physa1989,vinegar:spefe1991,tutunjian:la1992,morriss:la1997,zhang:spwla1998,latorraca:spwla1999,appel:spwla2000,freedman:spe2001,lo:SPE2002,zhang:spwla2002,bryan:jcpt2003,freedman:spe2003,hirasaki:mri2003,chen:spe2004,freedman:petro2004,bryan:spe2005,winkler:petro2005,mutina:amr2005,nicot:spwla2006,straley:spwla2006,freed:jcp2007,nicot:spwla2007,burcaw:spwla2008,mutina:jpca2008,yang:jmr2008,hurlimann:petro2009,lisitza:ef2009,kantzas:jcpt2009,zielinski:lang2010,zielinski:ef2011,yang:petro2012,jones:spe2014,chen:cpc2014,freedman:rsi2014,benamsili:ef2014,stapf:ef2014,korb:jpcc2015,vorapalawut:ef2015,jones:acis2015,ordikhani:ef2016,shikhov:amr2016,singer:SPWLA2017,singer:EF2018,kausik:petro2019,markovic:fuel2020}. However, the $^1$H NMR relaxation mechanism in crude oils at high viscosity such as heavy-oils and bitumen remains elusive and a topic of great debate. 

One possible NMR relaxation mechanism in crude oils is surface paramagnetism \cite{chen:cpc2014,benamsili:ef2014,korb:jpcc2015,vorapalawut:ef2015,ordikhani:ef2016}, whereby the maltenes in the crude oils diffuse in and out of the asphaltene macro-aggregates, during which time they come into contact with the paramagnetic sites on the asphaltene surface. Another possible relaxation mechanism in crude oils is enhanced $^1$H-$^1$H dipole-dipole relaxation \cite{zega:physa1989,vinegar:spefe1991,tutunjian:la1992,morriss:la1997,zhang:spwla1998,lo:SPE2002,hirasaki:mri2003,yang:jmr2008,zielinski:ef2011,yang:petro2012,singer:EF2018,kausik:petro2019}, whereby the relaxation of the maltenes is enhanced by confinement from the transient nano-pores of the asphaltene macro-aggregates. Similarly, $^1$H-$^1$H dipole-dipole relaxation is also postulated to dominate for light hydrocarbons in the organic nano-pores of kerogen \cite{washburn:cmr2014,singer:petro2016,fleury:jpse2016,zhang:geo2017,washburn:jmr2017,tandon:spwla2019,xie:spwla2019,parambathu:arxiv2020}. In fact, crossed-linked asphaltenes have been shown to be a good model for kerogen when modeling of the equilibrium partitioning of hydrocarbons in nanoporous kerogen particles \cite{liu:EF2019}.

In order to investigate the NMR relaxation mechanism in heavy-oils and bitumen, we previously reported a series of NMR measurements on polydisperse polymers and polymer-heptane mixes \cite{singer:SPWLA2017,singer:EF2018}. Polymers are known to have similar rheology as heavy oils \cite{abivin:ef2012}, making them good models for studying the rheology of viscous fluids. These polymers also have negligible amounts of paramagnetics impurities ($<$ 100 ppm according to EPR), which makes them a good model for studying $^1$H-$^1$H dipole-dipole relaxation with measurements and molecular dynamics (MD) simulations. Many studies have been reported on the $^1$H-$^1$H dipole-dipole relaxation of monodisperse polymers, including field-cycling relaxometry \cite{kariyo:prl2006,kariyo:macro2008b,kruk:pnmrs2012} and multiple-quantum techniques \cite{graf:prl1998,chavez:macro2011,chavez:macro2011b,mordvinkin:jcp2017}, from which a wealth of information about the molecular dynamics of monodisperse polymers is obtained. In our case, we use polydisperse polymers since bitumen and heavy-oils are highly polydisperse fluids. Furthermore, the polydisperse polymers are viscosity standards designed to have minimal shear-rate dependence on viscosity, which is important when comparing with NMR relaxation which is measured at zero shear-rate. 

We previously showed that at high viscosities, the log-mean relaxation time $T_{1LM}$ for the polydisperse polymers becomes independent of viscosity and proportional to frequency $T_{1LM} \propto f_0$ \cite{singer:SPWLA2017,singer:EF2018}. This behavior presents significant deviations from the traditional Bloembergen, Purcell and Pound (BPP) model for $^1$H-$^1$H dipole-dipole relaxation of monodisperse hard-spheres \cite{bloembergen:pr1948} where $T_{1,{\rm BPP}} \propto f_0^2 \eta/T$ is predicted at high viscosities. Furthermore for the polydisperse polymers, we find that the ``plateau" value normalized to $f_0$ = 2.3 MHz, $T_{1LM}\times 2.3/\!f_0 \simeq$ 3 ms, is the {\it same} as previously reported bitumen data. This implies that the relaxation mechanism is independent of the paramagnetic concentration, and therefore that $^1$H-$^1$H dipole-dipole relaxation dominates over paramagnetism at high viscosities. 

This then lead us to develop a phenomenological model based on $^1$H-$^1$H dipole-dipole relaxation which accounts for $T_{1LM}$ plateau at high viscosities by lowering the frequency exponent in the BPP model \cite{singer:SPWLA2017,singer:EF2018}. Lowering the frequency exponent implies a distribution in molecular correlation times of the viscous fluid, which is a similar approach to the phenomenological Cole-Davidson function \cite{davidson:jcp1951,lindesy:jcp1980} commonly used for dielectric and NMR data of glycerol \cite{flamig:jpcb2020} and monodisperse polymers \cite{kruk:pnmrs2012}. Our model also includes the presence of internal motions of the polymer branches through the Lipari-Szabo model \cite{lipari:jacs1982,lipari:jacs1982b}.

In this study, we further test our model on polydisperse polymers using $T_{1}$ field cycling relaxometry and $T_{1\rho}$ relaxation in the rotating frame. In the absence of paramagnetic impurities, we report on the anomalous viscosity dependence $T_{2LM}\propto (\eta/T)^{-1/2}$ for the polydisperse polymers at high viscosity, where a similar anomalous behavior was previously reported for bitumen \cite{yang:petro2012,kausik:petro2019}. This again presents a significant departure from BPP where $T_{2,{\rm BPP}} \propto (\eta/T)^{-1}$ is predicted at high viscosities. We report on MD simulations of $T_1$ and $T_2$ by $^1$H-$^1$H dipole-dipole relaxation of the polymer with viscosities in the range $\eta = 1 \leftrightarrow $ 1,000 cP. The MD simulations show that $T_{1} \propto f_0$ at high frequencies ($f_0 \gtrsim 100 $ MHz), specifically $T_{1}\times 2.3/\!f_0 \simeq$ 3 ms, in good agreement with measurements and our model at high viscosity. The MD simulations also confirm the dominance of int{\it ra}-molecular over int{\it er}-molecular $^1$H-$^1$H relaxation at high viscosity, which was previously only assumed to be the case.

\section{Methodology}\label{sc:Methodology}

\subsection{Experimental}\label{subsc:Experimental}

\begin{figure}[!ht]
	\begin{center}
\includegraphics[width=1\columnwidth]{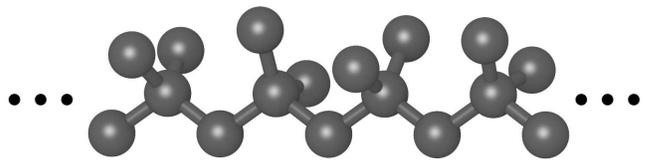} 
	\end{center}
	\caption{Illustration of poly(isobutene), where only carbon atoms are shown for clarity. MD simulations of poly(isobutene) were performed with a 16-mer (i.e. 64 carbon atoms) with $M_w$ = 912 g/mol and $\eta \simeq $ 1,000 cP at ambient.} \label{fg:poly}
\end{figure}

The polymers used in this study are listed in Table \ref{tb:Brookfield}. The average molecular weight $M_w$ and poly-dispersivity index $M_w/M_n$ were measured using gel permeation chromatography (GPC) using an Agilent Technologies 1200 module. The data in Table \ref{tb:Brookfield} indicate that the polymers are highly dispersive, up to $M_w/M_n \simeq 3.11$ in the case of B360000 poly(isobutene). The large polydispersivity of the polymers make them ideal for comparing with crude-oils, which are also highly dispersed as evidenced by their wide $T_2$ distributions \cite{freedman:spe2001}. In the case of the three poly(isobutene) polymers in Table \ref{tb:Brookfield}, the viscosity fit well to the functional form $\eta \simeq A\, M_w^{\alpha}$ \cite{holden:japs1965}, with $\alpha \simeq 2.4 $ and $A \simeq 1.07 \times 10^{-4}$ in units of (cP) and (g/mol) at ambient \cite{singer:EF2018}. An illustration of a section of poly(isobutene) is shown in Fig. \ref{fg:poly}, which was used for molecular dynamics (MD) simulations.

\begin{table}[!ht]
	\centering
	\begin{tabular}{cccccc}				
		\hline
		Name$^{{\strut}{}}$	& Composition& 	$\eta$ (25$^{\circ}$C)& $\eta$ (40$^{\circ}$C)& $M_w$ & $\frac{M_w}{M_n}$\\
		$^{{\strut}{}}$	& & (cP)& (cP)	&	 (g/mol) & \\
		\hline
		B1060$^{{\strut}{}}$	 & 	Poly(1-decene)& 	1,040&  385 &  4,204 & 1.49\\
		B10200$^{{\strut}{}}$		& Poly(isobutene)& 	10,700& 4,060 &  2,256& 2.13\\
		B73000$^{{\strut}{}}$		& Poly(isobutene)& 	68,100& 28,700 & 4,368 & 2.53\\
		B360000	$^{{\strut}{}}$	& Poly(isobutene)& 	333,000& 	102,000 & 9,436 & 3.11
	\end{tabular}
	\caption{Brand name, composition, viscosity $\eta$ at 25$^{\circ}$C and 40$^{\circ}$C, average molecular weight $M_w$, and polydispersivity index $M_w/M_n$, for the Brookfield viscosity standards used in this study.} \label{tb:Brookfield}
\end{table}

The viscosity measurements were made using a Brookfield AMETEK viscometer. The viscosities did not depend on shear-rate (within experimental uncertainties), thereby making them suitable viscosity ``standards" for comparing with NMR measurements which are measured at zero shear-rate. The viscosity measurements were made at both ambient temperature ($\simeq$ 25 $^{\circ}$C) and at equilibrated temperatures of 40 $^{\circ}$C using a circulation heat-bath. The viscosity data at 40 $^{\circ}$C is used as a proxy for the NMR data at 38.4 $^{\circ}$C.

\begin{figure}[!ht]
	\begin{center}
\includegraphics[width=1\columnwidth]{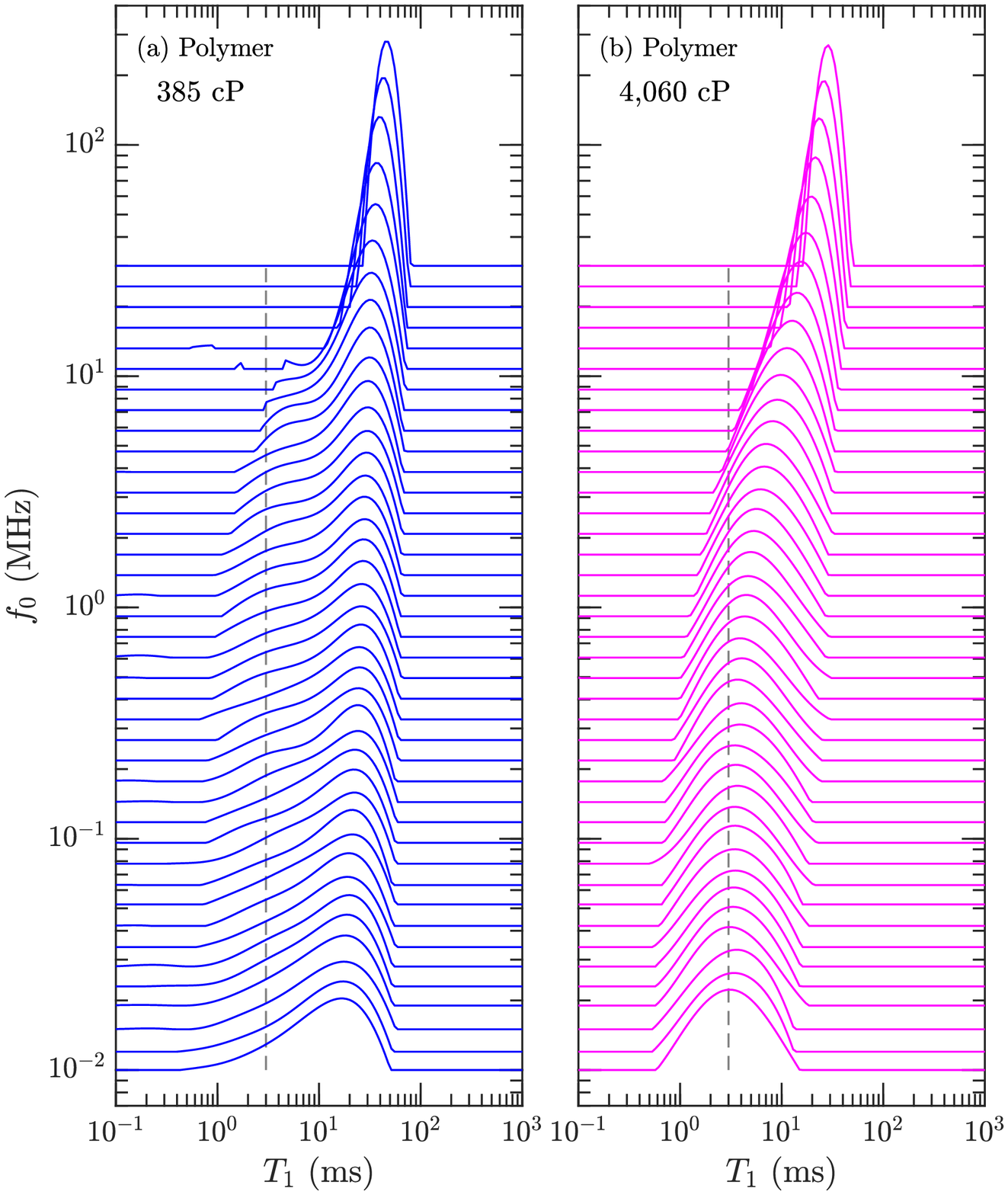} 
	\end{center}
	\caption{$T_{1}$ distributions from field cycling (FC) relaxometry at 38.4$^{\circ}$C for polymers listed in Table \ref{tb:Brookfield}. Dashed vertical line indicates FC ramp-time.} \label{fg:FC_T1_A}
\end{figure}

A 1 GHz electron paramagnetic resonance (EPR) apparatus was used to measure the concentration of paramagnetic ions {\it plus} the (weight equivalent) concentration of free radicals (i.e. unpaired valence electrons), which both contribute to NMR paramagnetic relaxation. The EPR data on the Brookfield viscosity standards indicated $< 100$ ppm paramagnetic impurities (i.e. the signal was below the detection limit of the apparatus). The paramagnetic concentration in the polymers is at least an order of magnitude less than the estimated $\simeq$ 1,000 ppm for Athabasca bitumen \cite{zhao:ef2007,singer:EF2018}.

\begin{figure}[!ht]
	\begin{center}
\includegraphics[width=1\columnwidth]{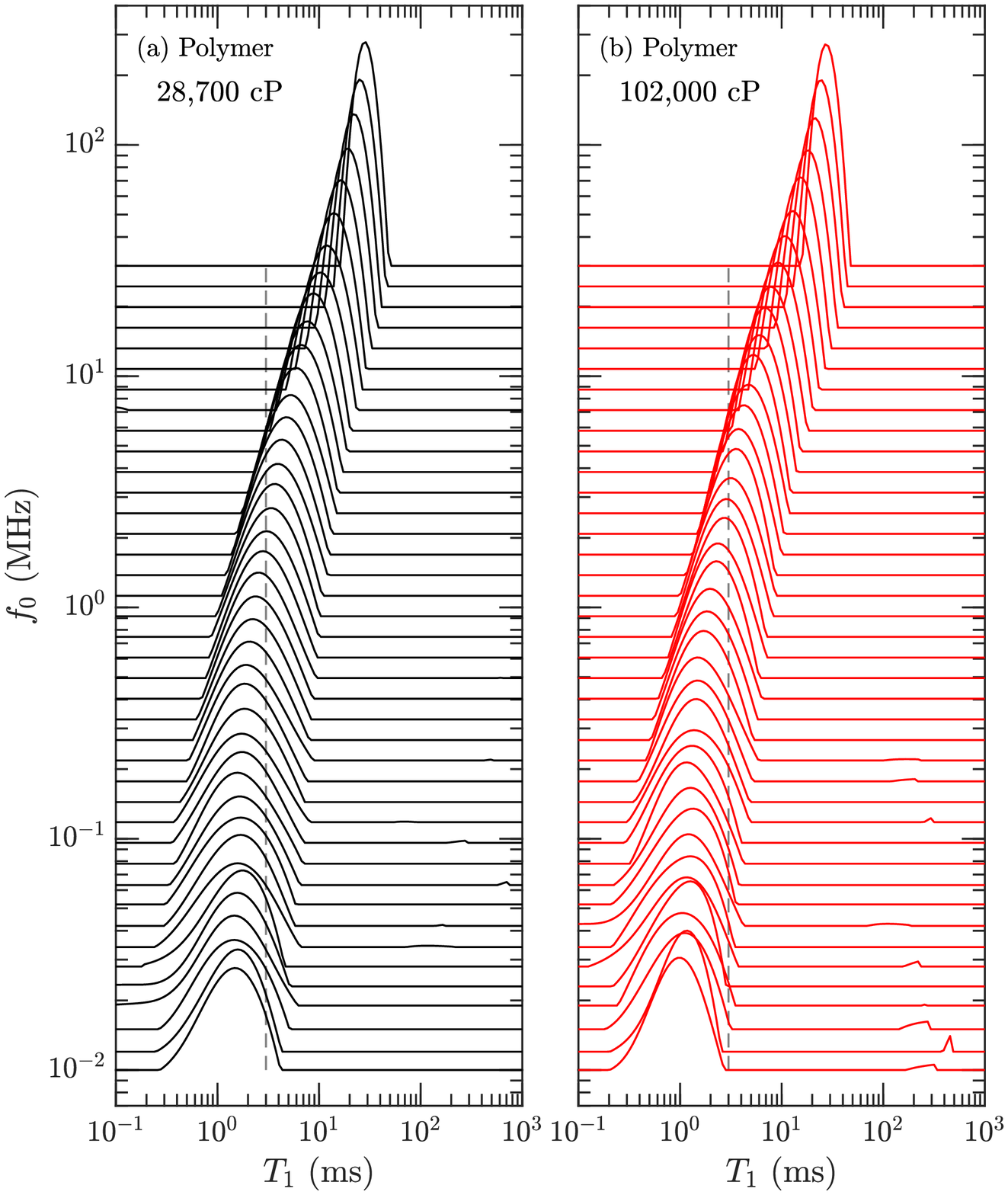} 
	\end{center}
	\caption{$T_{1}$ distributions from field cycling (FC) relaxometry at 38.4$^{\circ}$C for polymers listed in Table \ref{tb:Brookfield}. Dashed vertical line indicates FC ramp-time.} \label{fg:FC_T1_B}
\end{figure}

\begin{figure}[!ht]
	\begin{center}
\includegraphics[width=1\columnwidth]{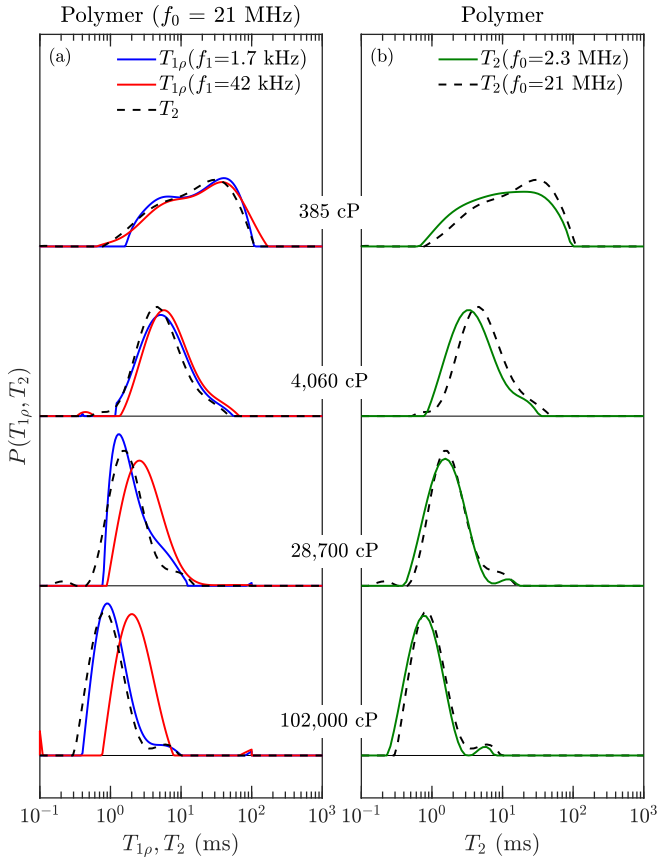}  
	\end{center}
	\caption{(a) $T_{1\rho}$ distributions at $f_1$ = 1.7 kHz and $f_1$ = 42 kHz, along with $T_2$ distributions, all measured at $f_0$ = 21 MHz and 38.4$^{\circ}$C. (b) $T_2$ distributions at $f_0$ = 2.3 MHz and 35$^{\circ}$C, along with $T_2$ distributions at $f_0$ = 21 MHz and 38.4$^{\circ}$C.} \label{fg:T1_rho}
\end{figure}

$^1$H NMR $T_1$ and $T_2$ measurements at a resonance frequency of $\omega_0/2\pi = f_0$ = 2.3 MHz were made with a GeoSpec2 from Oxford Instruments, with a 29 mm diameter probe. The samples were measured at ambient conditions ($\simeq$ 25$^{\circ}$C) and after temperature equilibration to $\simeq$ 30$^{\circ}$C. Additional measurements at $\simeq$ 35$^{\circ}$C were made by turning off the chiller and equilibrating to the magnet temperature. $T_1$ and $T_2$ measurements at $f_0$ = 22 MHz and $\simeq$ 30$^{\circ}$C were made with a special spectrometer from MR Cores at Core Laboratories, with a 30 mm diameter probe. $T_1$ and $T_2$ measurements at $f_0$ = 400 MHz at $\simeq$ 25$^{\circ}$C were made with a Bruker Avance spectrometer, in a 5 mm diameter probe. 

$^1$H NMR $T_1$, $T_{1\rho}$, and $T_2$ measurements at $f_0$ = 21 MHz and 38.4$^{\circ}$C were made at IFP-EN on a Maran Ultra, with a 18 mm diameter probe. The $T_{1\rho}$ measurements in the rotating frame \cite{kimmich:book,steiner:cpl2010} were made with a spin-locking frequency $\omega_1/2\pi = f_1 = $ 1.7 kHz $\leftrightarrow$ 42 kHz. Field cycling measurements were made on a Stelar fast field-cycling (FC) relaxometer from $f_0 = 0.01 \leftrightarrow$ 30 MHz at 38.4$^{\circ}$C, with a 10 mm diameter probe. All of the above $T_2$ measurements were measured using a CPMG sequence with echo spacings of $T_E = 0.1$ ms or less, except at $f_0$ = 400 MHz where $T_2$ was estimated from $T_2 \simeq 1/\pi\Delta \! f$, where $\Delta \! f$ is the width of the NMR spectrum. The hydrogen index ($HI \simeq$ 1.17) of the polymers is discussed in \cite{singer:EF2018}.

The $T_1$, $T_{1\rho}$, and $T_2$ distributions of the pure polymers were determined using inverse Laplace transforms \cite{venkataramanan:ieee2002,song:jmr2002}. The FC $T_{1}$ distributions shown in Figs. \ref{fg:FC_T1_A} and \ref{fg:FC_T1_B} tend to narrow with increasing frequency due to larger (absolute) longitudinal cross-relaxation \cite{kalk:jmr1976,kowalewski:book} (a.k.a. spin-diffusion). Figs. \ref{fg:FC_T1_A} and \ref{fg:FC_T1_B} also show the finite ramp-time of 3 ms required to ramp the field up and down. While in theory this does not effect the $T_1$ acquisition \cite{kimmich:pnmrs2004}, it has been noted that it does effect broad $T_1$ distributions with fast relaxing components \cite{roos:jbio2015,ward:jpc2018}. Our relaxation model indicates that this is likely the case for the 102,000 cP polymer, where the log-mean $T_{1LM}$ is most likely overestimated by a factor $\simeq 2^{1/2}$ at low frequencies.

The $T_{1\rho}$ distributions at $f_1$ = 1.7 kHz and $f_1$ = 42 kHz are shown in Fig. \ref{fg:T1_rho}, alongside the $T_2$ distribution. $T_{1\rho}$ and $T_2$ distributions tend to narrow with increasing viscosity, which is opposite to the trend in polydispersivity index $M_w/M_n$ in Table \ref{tb:Brookfield}. The narrowing may therefore be a result of larger transverse cross-relaxation (in the low frequency limit) with increasing viscosity (i.e. increasing correlation time) \cite{kowalewski:book}. In the case of the most viscous polymer, Fig. \ref{fg:T1_rho} shows that $T_{1\rho}$ increases when going from $f_1$ = 1.7 kHz to 42 kHz, indicating the presence of molecular correlation times shorter than $\tau \lesssim (2 \omega_1)^{-1} \simeq$ 1 $\mu$s.

The log-mean values $T_{1LM}$, $T_{1\rho LM}$ and $T_{2LM}$ of the distributions are used for data analysis and fitting to the model, where for example $T_{1LM} = \exp\! \left<{\rm ln}\,T_{1}\right>$, which is justified from the constituent viscosity model \cite{freedman:spe2001}. As shown in the Supporting Information, the effects of dissolved oxygen on $T_{1}$ as a function of frequency were measured for $n$-heptane, and the concentration of dissolved oxygen $C_{\rm O_2}$ in the polymer-heptane mix was predicted by MD simulations. The results indicate that the effects of dissolved oxygen on $T_{1}$ (and $T_2$) are negligible for all the polymers at all frequencies.

\subsection{Relaxation model}\label{subsc:Models}

The underlying expressions for $T_1$, $T_2$ and $T_{1\rho}$ in an isotropic system are given by \cite{mcconnell:book,cowan:book}:
\begin{align}
\frac{1}{T_{1}} &= J(\omega_0) + 4 J(2\omega_0), \nonumber \\
\frac{1}{T_{2}} &= \frac{3}{2} J(0) + \frac{5}{2} J(\omega_0) + J(2\omega_0), \label{eq:T12J}  \\
\frac{1}{T_{1\rho}} &= \frac{3}{2} J(2\omega_1) + \frac{5}{2} J(\omega_0) + J(2\omega_0). \nonumber 
\end{align}
$J(\omega_0)$ is the spectral density at the resonance frequency $\omega_0 = 2\pi f_0$. The expression for $T_1$ in the rotating frame, $T_{1\rho}$, is similar to $T_2$ except that the zero-frequency term $J(0)$ is replaced by $J(2\omega_1)$ where $\omega_1 = 2\pi f_1 $ is the spin-locking frequency \cite{kimmich:book}. Note that Eq. \ref{eq:T12J} does not assume a model for the spectral density $J(\omega)$.

\subsubsection{BPP model}\label{subsubsc:BPP}

The BPP model for the spectral-density $J_{\rm BPP}(\omega)$ for intra-molecular $^1$H-$^1$H dipole-dipole relaxation is given by the following \cite{bloembergen:pr1948}:
\begin{align}
J_{\rm BPP}(\omega) &= \frac{1}{3} \Delta\omega_R^2  \frac{2\tau_{\eta}}{1+\left(\omega\tau_{\eta}\right)^2}, \label{eq:BPP1}\\
\tau_{\eta} &= \frac{4\pi }{3k_B} R^3 \frac{\eta}{T}, \label{eq:BPP2}\\
\Delta\omega_R^2 &=  \frac{9}{20} \! \left(\frac{\mu_0}{4\pi}\right)^{\! 2} \! \hbar^2 \gamma^4 \frac{1}{N}\sum\limits_{i \neq j}^{N}{\frac{1}{r_{ij}^6}}. \label{eq:BPP3}
\end{align}
The BPP model assumes the Stokes-Einstein-Debye relation for hard spheres, where $\tau_{\eta}$ is the rotational correlation-time, $\eta/T$ is viscosity over temperature, and $R$ is the Stokes radius. The constant $\Delta\omega_R^2 $ is the ``second-moment" (i.e. the strength) of the intra-molecular $^1$H-$^1$H dipole-dipole interactions (where $r_{ij}$ is the $^1$H-$^1$H distance between pairs $i$ and $j$). Note that the BPP model is only valid in the motional-narrowing regime where $\Delta \omega_R \tau_{\eta} \ll 1$ \cite{cowan:book}, which is assumed to be the case throughout. 

The BPP model introduces the important concept of the  fast-motion (i.e. low viscosity) regime ($\omega_0 \tau_{\eta} \ll 0.615$) where $T_1/T_2 = 1$:
\begin{align}
T_{1,{\rm BPP}} = T_{2,{\rm BPP}} &\propto \left(\frac{\eta}{T}\right)^{-1} \,\,\, {\rm for }\,\,\, \omega_0 \tau_{\eta} \ll 0.615 \label{eq:BPP<}
\end{align}
and the slow-motion (i.e. high viscosity) regime ($\omega_0 \tau_{\eta} \gg 0.615$) where $T_1/T_2 > 1$:
\begin{align}
T_{1,{\rm BPP}} &\propto f_0^2\frac{\eta}{T} \qquad \,\,{\rm for } \,\,\, \omega_0 \tau_{\eta} \gg 0.615 \label{eq:BPP>T1} \\
T_{2,{\rm BPP}} &\propto \left(\frac{\eta}{T}\right)^{-1} \quad {\rm for } \,\,\,\omega_0 \tau_{\eta} \gg 0.615. \label{eq:BPP>T2} 
\end{align}

\subsubsection{New relaxation model}\label{subsubsc:NewModel}

The BPP model fails for polydisperse polymers and bitumen at high-viscosity. In particular, BPP predicts that $T_1 \propto f_0^2\eta/T$ at high-viscosities, while measurements clearly indicate that $T_{1LM}\propto f_0$ is independent of viscosity. As such, a phenomenological model was developed where the frequency exponent in Eq. \ref{eq:BPP1} is lowered from the BPP value $(\omega\tau)^2$ to $(\omega\tau)^1$. This has the effect of dropping $\tau_{\eta}$ (and therefore $\eta/T$) out of the equation in the slow-motion regime ($\omega_0\tau_{\eta} \gg 1$) \cite{singer:SPWLA2017,singer:EF2018}. In the Supporting Information, we show that our model for the frequency exponent $(\omega\tau)^1$ is similar to the limiting case of the phenomenological Cole-Davidson function commonly used for dielectric data for glycerol \cite{davidson:jcp1951,lindesy:jcp1980}, as well as for NMR data of glycerol and monodisperse polymers \cite{kruk:pnmrs2012}.

The consequence of changing the frequency exponent is to introduce a distribution $P_R(\tau)$ in local rotational correlation times $\tau$. This is justified by Woessner's theories which show that as the molecule becomes less spherical, the internal motions in the molecule become more complex, and the distribution in correlation times becomes more pronounced \cite{woessner:jcp1962,woessner:jcp1965}. We also note that according to Woessner's theories, simple fluids show a large distribution in correlation times when their motion is restricted by nano-confinement \cite{orazio:prb1990}.

Besides changing the frequency exponent, our model also allows for the existence of internal motions of non-rigid polymers using the Lipari-Szabo (LS) model \cite{lipari:jacs1982,lipari:jacs1982b}. Changing the frequency exponent to $(\omega\tau_{\eta})^1$ in the BPP model and applying the LS model results in the following spectral density:
\begin{align}
J_P(\omega)  &=\frac{1}{3} \Delta\omega_R^2  \left( S^2 \frac{2\tau_R}{1+\omega\tau_R} + \left(1-S^2\right) \frac{2\tau_L}{1+\omega\tau_L} \right)\label{eq:NewLS}
\end{align}
\noindent where the subscript $P$ in $J_P(\omega)$ refers to the ``Plateau", and $\tau_R \gg \tau_L$ is assumed. $\tau_R$ is defined as the slow rotational correlation-time of the whole polymer molecule, which depends on viscosity. The order parameter $S^2$ is a measure of the rigidity of the polymer molecule, where $S^2$ = 1 for completely rigid molecules with no internal motion of the polymer branches, and $S^2$ = 0 for completely non-rigid molecules with full internal motion of the polymer branches. $\tau_L$ is the local correlation-time, which characterizes the fast $\tau_L$ ($\simeq$ 10's ps) motions of the polymer branches. 

Eq. \ref{eq:NewLS} predicts the following expression in the slow-motion ($\omega\tau_R \gg 1$) regime:
\begin{align}
J_{P}(\omega)  &\simeq \frac{1}{3} \Delta\omega_R^2  \left( S^2 \frac{2}{\omega} + \left(1-S^2\right) 2\tau_L + \dots\right) \label{eq:NewLSslow}
\end{align}
which leads to the following approximation for $T_{1LM}$:
\begin{align}
\frac{1}{T_{1LM}} &\simeq \frac{2 \Delta\omega_R^2 S^2}{\omega_0}\left(1+\frac{5}{3}\frac{1-S^2}{S^2}\omega_0\tau_L  + \dots\right). \label{eq:LS_T1}
\end{align}
In other words, the leading order term is $T_{1LM} \propto f_0$, which is independent of viscosity. A deviation from linearity $T_{1LM} \propto f_0$ occurs at high frequencies when  $\omega_0\tau_L \simeq 1$. This turns out to be more prominent for bitumen than for the polydisperse polymers, where $\tau_L$ is larger for bitumen (see below). We note that the temperature dependence of $\tau_L$ is most likely present but much less than the temperature dependence of $\tau_R$ (which depends on viscosity).

Eq. \ref{eq:NewLS} also predicts the following approximation for $T_{2LM}$ in the slow-motion regime:
\begin{align}
\frac{1}{T_{2LM}} &\simeq \frac{10}{3}\Delta\omega_R^2 S^2 \tau_R + \dots,  \label{eq:LS_T2}\\
{\rm where } \,\,\,\tau_R &= \left(\tau_\eta \tau_0\right)^{1/2}. \nonumber
\end{align}
The phenomenological relation $\tau_R = \left(\tau_\eta \tau_0\right)^{1/2}$ is introduced in \cite{korb:jpcc2015,kausik:petro2019}, which relates $\tau_R$ to the Stokes-Einstein-Debye correlation time $\tau_{\eta}\propto \eta/T$ (Eq. \ref{eq:BPP2}) at high viscosities. $\tau_0$ is a constant, which leads to the prediction that $T_{2LM} \propto \tau_R^{-1} \propto (\eta/T)^{-1/2}$, in agreement with previously published bitumen data \cite{yang:petro2012,kausik:petro2019} and the polydisperse polymer data shown below. 

Two theories have been proposed for the relation $T_{2LM} \propto (\tau_{\eta} \tau_0)^{-1/2} \propto (\eta/T)^{-1/2}$ at high viscosity. The first theory by Korb {\it et al.} \cite{korb:jpcc2015} stipulates that $\tau_0$ (referred to as $\tau_{1D}$ in \cite{korb:jpcc2015}) corresponds to a quasi-1D translational diffusion time of a maltene molecule within the transient nano-porous network of quasi immobile asphaltene macro-aggregates.
$\tau_0 \propto \eta/T$ below a critical viscosity $\eta \ll \eta_c$ (with $\eta_c \simeq$ 300 cP), while $\tau_0$ is constant above the critical viscosity $\eta \gg \eta_c$. While \cite{korb:jpcc2015} uses this theory in the context of paramagnetism, their model for $\tau_R$ can also apply here in the context of $^1$H-$^1$H dipole-dipole relaxation. The second theory \cite{kausik:petro2019} arrives at the same dependence of $T_{2LM} \propto (\tau_{\eta} \tau_0)^{-1/2} \propto (\eta/T)^{-1/2}$ at high viscosity, but $\tau_0$ (referred to as $\tau_a$ in \cite{kausik:petro2019}) is dominated by the maltene's residence time $\tau_{res}$ in the asphaltene cluster (i.e. $ \tau_0 \simeq \tau_{res}$), which is independent of viscosity.

Finally we note that the expression for $T_{2LM}$ (Eq. \ref{eq:LS_T2}) contains the factor 10/3, implying that $T_{2}$ is in the fast-motion regime (i.e. independent of frequency) even when $\omega_0\tau_R \gg 1$. This is motivated by the interpretation of the polydisperse polymer and bitumen data presented below.

\subsection{MD simulations}\label{subsc:MDsimulations}
 
 \begin{figure}[!ht]
 	\begin{center}
 \includegraphics[width=1\columnwidth]{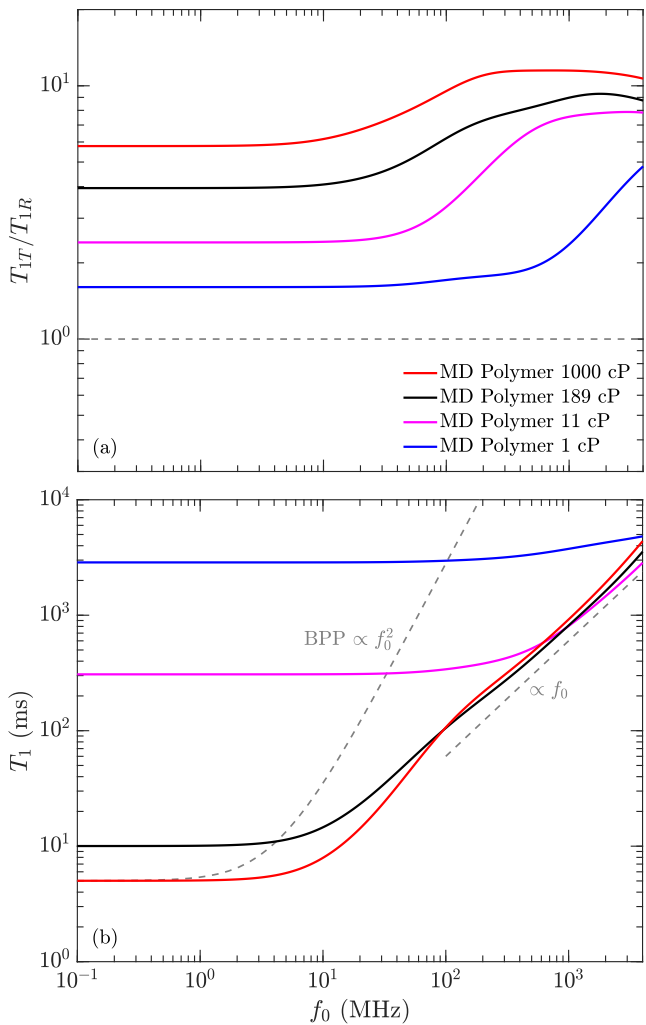}			
 	\end{center}
 	\caption{(a) MD simulations of the ratio of inter-molecular to intra-molecular relaxation times $T_{1T}/T_{1R}$ as a function of frequency for the four polymers. (b) Total relaxation time $T_1$ as a function of frequency for the four polymers. Also shown are the BPP prediction ($T_1 \propto f_0^2$), and the observed trend ($T_1 \propto f_0$) above $f_0 > $ 100 MHz.} \label{fg:MD_T1}
 \end{figure}
 
Molecular dynamics (MD) simulations of monodisperse polymers were conducted. Four different chain lengths of poly(isobutene) (see Fig. \ref{fg:poly}) were simulated at 25$^{\circ}$C: a 16-mer, 8-mer, 4-mer and 2-mer. The viscosity of the 16-mer and 8-mer were estimated using the relation $\eta = A\, M_w^{\alpha}$ \cite{holden:japs1965}, with $\alpha \simeq 2.4 $ and $A \simeq 1.07 \times 10^{-4}$ in units of (cP) and (g/mol) at ambient \cite{,singer:EF2018}. For example, in the case of the 16-mer where $M_w = 912$ g/mol, this predicts a viscosity of $\eta \simeq $ 1,000 cP. The viscosities of the 4-mer and 2-mer were predicted using the empirical relation $T_{2LM} \simeq 9.56 \left(\eta/T\right)^{-1}$ \cite{lo:SPE2002} in units of (ms) and (cP/K).

MD simulations of the intra-molecular ($T_{1R,2R}$) \cite{bloembergen:pr1948} and inter-molecular ($T_{1T,2T}$) \cite{torrey:pr1953,hwang:JCP1975} $^1$H-$^1$H dipole-dipole relaxation were then computed for the polymers, from which the total relaxation times ($T_{1,2}$) are calculated:
\begin{align}
\frac{1}{T_{1,2}} &= \frac{1}{T_{1R,2R}} + \frac{1}{T_{1T,2T}}. \label{eq:T1_RT}
\end{align}
The procedure for the MD simulations are the same as reported elsewhere \cite{singer:jmr2017,asthagiri:seg2018,singer:jcp2018}, and details are given in the Supporting Information.

Fig. \ref{fg:MD_T1}(a) shows the ratio of inter-molecular to intra-molecular relaxation times $T_{1T}/T_{1R}$ as a function of frequency for the four polymers. A value larger than unity $T_{1T}/T_{1R} >1$ indicates that intra-molecular relaxation dominates, while $T_{1T}/T_{1R} <1$ indicates that inter-molecular relaxation dominates. We find that $T_{1T}/T_{1R} > 1 $ increases with increasing viscosity, implying that intra-molecular relaxation dominates (by at least an order of magnitude) at high viscosities ($\eta \gtrsim$ 1,000 cP). $T_{2T}/T_{2R}$ (not shown) show similar results. These findings justify the assumption in our model (Eq. \ref{eq:NewLS}) that intra-molecular relaxation dominates over inter-molecular relaxation. 

Fig \ref{fg:MD_T1}(b) shows the total relaxation $T_{1}$ (Eq. \ref{eq:T1_RT}) as a function of frequency for the four polymers. The lowest viscosity (1 cP) polymer shows high $T_1$ values and minimal dispersion (i.e. minimal frequency dependence). On the other hand, the highest viscosity polymers show significant dispersion. The 189 cP and 1,000 cP polymers merge above $f_0 > $ 100 MHz into a linear relation $T_1 \propto f_0$. This behavior is exactly predicted by the model (Eq. \ref{eq:LS_T1}), namely that $T_1 \propto f_0$ is independent of viscosity. The $f_0 > $ 100 MHz region for the 1,000 cP polymer is compared with measurements and the model below.

We note that a similar relation was previously reported from MD simulations of heptane confined in a polymer matrix, where the surface relaxation $T_{1S}$ of heptane followed $T_{1S} \propto f_0$ under confinement \cite{parambathu:arxiv2020}. This implies a connection between the molecular dynamics of high-viscosity fluids and low-viscosity fluids under confinement. 

\section{Results and Discussions}\label{sc:Results}

The results and interpretation are organized as follows. In section A we present the $T_{1LM}$ data for polydisperse polymers and bitumen in the slow-motion (i.e. high-viscosity) regime, and we use Eq. \ref{eq:LS_T1} to extract the free parameters $S^2$ and $\tau_L$ (Table \ref{tb:SlowMotion}). In section B we present the $T_{2LM}$ data for polydisperse polymers and bitumen in the slow-motion (i.e. high-viscosity) regime, and we use Eq. \ref{eq:LS_T2} to extract the free parameter $\tau_0$ (Table \ref{tb:SlowMotion}). In section C we present the full frequency dependence of $T_{1LM}$ and $T_{2LM}$ data for the polydisperse polymers spanning both the fast- and slow-motion regimes, and we use the full expression Eq. \ref{eq:NewLS} to extract $\tau_0$ for each polymer (Table \ref{tb:FastMotion}). 

\subsection{$T_{1LM}$ for polydisperse polymers and bitumen}\label{subsc:T1}

\begin{figure}[!ht]
	\begin{center}
\includegraphics[width=1\columnwidth]{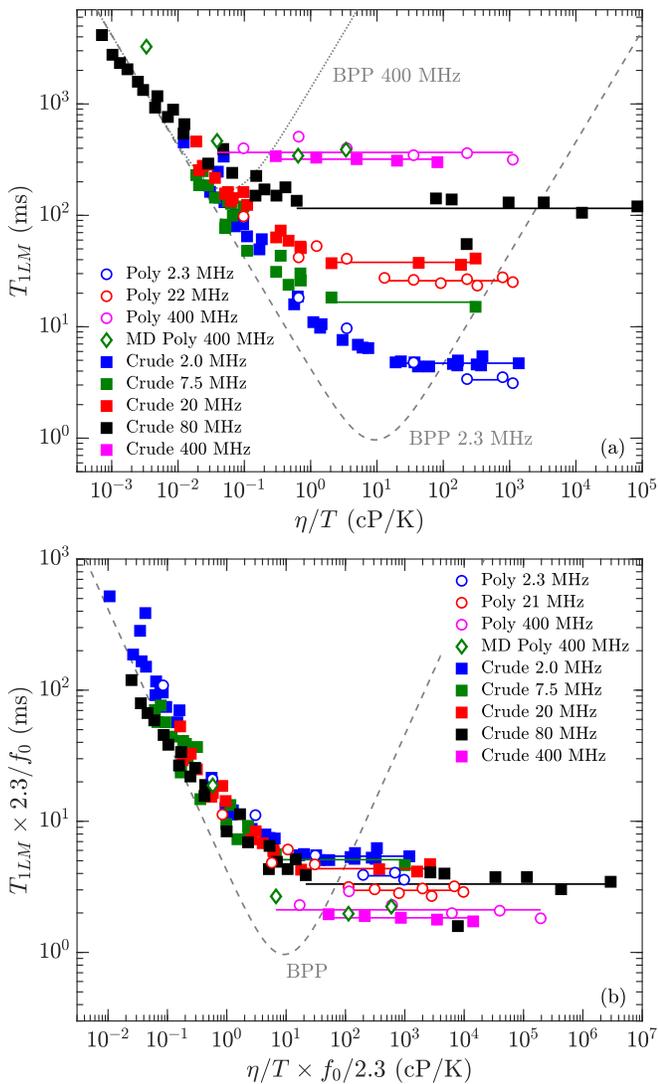} 
	\end{center}
	\caption{(a) $T_{1LM}$ vs. $\eta/T$ for the polydisperse polymers in Table \ref{tb:Brookfield} at 2.3 MHz, 21 MHz, and 400 MHz, and MD simulations of the polymer poly(isobutene) in Fig. \ref{fg:poly} at 400 MHz. Also shown are previously published crude-oil and bitumen data at 2.0 MHz (LaTorraca {\it et al.} \cite{latorraca:spwla1999}, Yang {\it et al.} \cite{yang:jmr2008}), at 7.5 MHz and 20 MHz (Zhang {\it et al.} \cite{zhang:spwla2002}), at 80 MHz (Vinegar {\it et al.} \cite{vinegar:spefe1991}), and at 400 MHz (Kausik {\it et al.} \cite{kausik:petro2019}). Horizontal lines at each frequency indicate log-average of $T_{1LM}$ in slow-motion (i.e. high-viscosity) region. Curved lines are the BPP prediction \cite{bloembergen:pr1948} at 2.3 MHz and 400 MHz. (b) Same data as in (a), but plotted on frequency normalized axes (normalized to $f_0$ = 2.3 MHz).} \label{fg:T1crude}
\end{figure}

The results for $T_{1LM}$ for crude oils (including bitumen) and polydisperse polymers are shown in Fig. \ref{fg:T1crude}(a). The crude-oil data are taken from various sources listed in the legend. The most recent addition is the bitumen data at $f_0 = $ 400 MHz by Kausik {\it et al.} \cite{kausik:petro2019}, measured over a range of temperatures (30$^{\circ}$C $\leftrightarrow$ 90$^{\circ}$C). 

Also shown in Fig. \ref{fg:T1crude}(a) is the BPP prediction \cite{bloembergen:pr1948} at $f_0$ = 2.3 MHz and 400 MHz from Eq. \ref{eq:BPP1}. The crude oils roughly follow the BPP prediction $T_{1LM} \propto(\eta/T)^{-1}$ at low viscosities, however $T_{1LM}$ clearly plateaus at high viscosity. Also shown are the MD simulations at 400 MHz for the polymers, which are consistent with the polydisperse polymer measurements.

Fig. \ref{fg:T1crude}(b) shows the same data as Fig. \ref{fg:T1crude}(a) but on a frequency normalized scale. More specifically, the $x$-axis ($\eta/T$) is multiplied by $f_0/2.3$ with $f_0$ in units of MHz, while the $y$-axis ($T_{1LM}$) is divided by $f_0/2.3$. Frequency normalizing has the effect of collapsing the frequency dependence of the BPP model onto one universal curve \cite{zhang:spwla2002}. It also has the effect of collapsing the bitumen and polymer data in the slow-motion regime onto one plateau value given by $T_{1LM}\times 2.3/f_0 \simeq $ 3 ms, i.e. $T_{1LM}\propto f_0$. The more recent bitumen data at 400 MHz shows a slight departure from the low frequency data, namely the plateau value is lower than at lower frequencies. As shown below, the model takes this departure into account with the $\omega_0\tau_L$ term in Eq. \ref{eq:LS_T1}.

\begin{figure}[!ht]
	\begin{center}
\includegraphics[width=1\columnwidth]{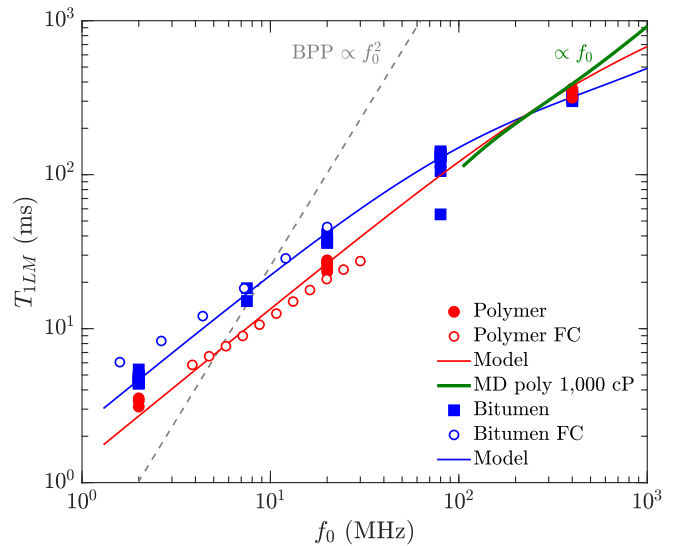} 
	\end{center}
	\caption{$T_{1LM}$ vs. $f_0$ data for polydisperse polymers and bitumen taken from Fig. \ref{fg:T1crude}(a), where only data from the slow-motion (i.e. high viscosity) regime are included. Bitumen data are taken from various sources (see Fig. \ref{fg:T1crude} caption), while field cycling (FC) data for bitumen are taken from \cite{kausik:petro2019}. Also shown are FC data for the 102,000 cP polymer. Solid curves are fits using the model in Eq. \ref{eq:LS_T1} with fitting parameters shown in Table \ref{tb:SlowMotion}. 
	MD simulations of the 1,000 cP polymer are shown above $f_0>$ 100 MHz, corresponding viscosity independent region (see Fig. \ref{fg:MD_T1}) where $T_1\propto f_0$. BPP prediction $T_1\propto f_0^2$ is also shown.} \label{fg:Omega}
\end{figure}

Fig. \ref{fg:Omega} shows $T_{1LM}$ data for the polydisperse polymers and the bitumen in the slow-motion regime (i.e. high-viscosity) regime, which corresponds to data within the horizontal lines (the log-average) in Fig. \ref{fg:T1crude}(a). The best fit to the new model using Eq. \ref{eq:NewLSslow} and Eqs. \ref{eq:T12J} are shown for both polydisperse polymers and bitumen, and the best fit parameters are shown in Table \ref{tb:SlowMotion}. The second moment is fixed to $\Delta\omega_R/2\pi$ = 20.0 kHz, which is the value for $n$-heptane \cite{singer:EF2018}. A Stokes radius of $R = $ 1.85 {\rm \AA} is used for the polymers, which is the value needed to match the correlation $T_{1LM,2LM} = 9.56\,(\eta/T)^{-1}$ in the low-viscosity regime \cite{lo:SPE2002}. A slightly larger Stokes radius of $R = $ 2.47 {\rm \AA} is used, which is the value needed to match the correlation $T_{1LM,2LM} = 4.0\,(\eta/T)^{-1}$ in the low-viscosity regime \cite{freedman:spe2001}. 

We note that the fact that $T_{1LM,2LM}$ are consistent with a constant Stokes radius $R$ in the low-viscosity regime, i.e. $R$ is independent of molecular size (and therefore viscosity), clearly shows that $T_{1,2}$ are probes of the {\it local} molecular dynamics. This is in stark contrast to the radius of gyration $R_g$ which depends on the molecular size \cite{singer:jmr2017}.

The first free-parameter in the model is the order parameter $S^2$, which characterizes the rigidity of the molecule. $S^2 \simeq 0.147$ is found for the polydisperse polymers, which is consistent with previously reported data for monodisperse polymers at high molecular-weights $M_w > $ 4,000 g/mol  \cite{graf:prl1998,kariyo:prl2006,kariyo:macro2008b}. The fit to bitumen indicates a lower $S^2 \simeq 0.085$, implying a less-rigid molecule (i.e. more isotropic internal-motions). The second free-parameter is the local correlation-time $\tau_L$, which characterizes the fast $\tau_L$ ($\simeq$ 10's ps) motions of the molecular branches. The fit indicates a local correlation time of $\tau_L \simeq 20 $ ns for the polydisperse polymers, and a longer $\tau_L \simeq 53 $ ns for bitumen. We note that unlike bitumen, the fit for the polydisperse polymers is not very sensitive to $\tau_L $, therefore an upper bound $\tau_L \lesssim 20 $ ns may be more appropriate for the polydisperse polymers.

\begin{table}[!ht]
	\centering
		\begin{tabular}{ccc|ccc}				
		\hline
		$^{{\strut}{}}$	& $\Delta\omega_R/2\pi$ & $R$ & $S^2$ & 	$\tau_L$ & 	$\tau_0$  	 \\
		$^{{\strut}{}}$	& (kHz) & (${\rm \AA}$) &  &	(ps) & (ns)\\
		\hline
		Bitumen	& 20 & 2.47 $^{{\strut}{}}$&$^{{\strut}{}}$ 0.085 & 53 & 268 \\
		Polymer	& 20& 1.85 $^{{\strut}{}}$&$^{{\strut}{}}$ 0.147 & 20 & 42
	\end{tabular}
	\caption{Results of the fitting parameters to the model for polydisperse polymers and bitumen in the high viscosity regime. $S_2$ and $\tau_L$ are optimized from data in Fig. \ref{fg:Omega} using Eq. \ref{eq:LS_T1}, while $\tau_0$ is optimized from data in Fig. \ref{fg:T2crude} using Eq. \ref{eq:LS_T2}. $\Delta\omega_R$ and $R$ are fixed.} \label{tb:SlowMotion}
\end{table}

Also shown in Fig. \ref{fg:Omega} are the MD simulations of the 1,000 cP polymer above $f_0>$ 100 MHz, corresponding viscosity independent region (see Fig. \ref{fg:MD_T1}) where $T_1\propto f_0$. The agreement between simulation and data/model in Fig. \ref{fg:Omega} is remarkable given that only a monodisperse model of poly(isobutene) is used in the simulations. This suggests that the $T_1\propto f_0$ behavior is generic  at high-viscosities, and that  $^1$H-$^1$H dipole-dipole relaxation dominates over paramagnetism at high-viscosities.

\subsection{$T_{2LM}$ for polydisperse polymers and bitumen}\label{subsc:T2}

\begin{figure}[!ht]
	\begin{center}
\includegraphics[width=1\columnwidth]{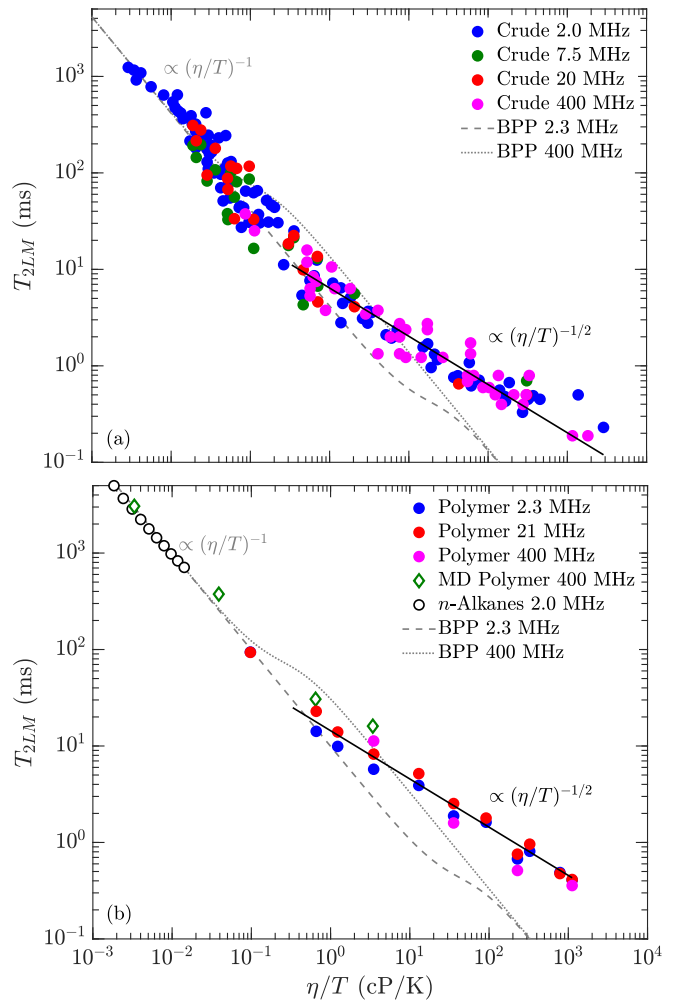}
	\end{center}
	\caption{(a) $T_{2LM}$ vs. $\eta/T$ for bitumen at 2.0 MHz (Vinegar {\it et al.} \cite{vinegar:spefe1991}, LaTorraca {\it et al.} \cite{latorraca:spwla1999}, Yang {\it et al.} \cite{yang:jmr2008}), at 7.5 MHz and 20 MHz (Zhang {\it et al.} \cite{zhang:spwla2002}), and at 400 MHz (Kausik {\it et al.} \cite{kausik:petro2019}). (b) $T_{2LM}$ vs. $\eta/T$ for the polydispersed polymers in Table \ref{tb:Brookfield} at 2.3 MHz, 21 MHz, and 400 MHz, and MD simulations of the polymers at 400 MHz. $n$-Alkane de-oxygenated data at 2.0 MHz (Shikov {\it et al.} \cite{shikhov:amr2016}). Solid lines are fits using the model in Eq. \ref{eq:LS_T2} for the region $\eta/T \gtrsim $ 0.3 cP/K (or $\eta \gtrsim $ 100 cP at ambient, equivalently), with fitting parameters listed in Table \ref{tb:SlowMotion}. Curved lines are the BPP prediction \cite{bloembergen:pr1948} at 2.3 MHz and 400 MHz.} \label{fg:T2crude}
\end{figure}

The results for $T_{2LM}$ for the crude oils are shown in Fig. \ref{fg:T2crude}(a), taken from various sources listed in the legend, with the most recent addition is the bitumen data at $f_0 = $ 400 MHz by Kausik {\it et al.} \cite{kausik:petro2019}. $T_{2LM}$ for the polydisperse polymers are shown in Fig. \ref{fg:T1crude}(b), along with de-oxygenated $n$-alkane data \cite{shikhov:amr2016}, and the MD simulation results for the polymers at 400 MHz.  

Solid lines are fits using the model in Eq. \ref{eq:LS_T2} with fitting parameters shown in Table \ref{tb:SlowMotion}, restricted to the high-viscosity region $\eta/T > 0.3 $ cP/K, or $\eta > 100 $ cP at ambient equivalently. Both the polydisperse polymers and bitumen data are consistent with $T_{2LM} \propto \left(\eta/T\right)^{-1/2}$ for viscosities higher than $\eta/T > 0.3 $ cP/K. The model indicates that $\tau_0$ is a factor $\simeq$ 6 larger for bitumen than for the polydisperse polymers. As discussed in Section \ref{subsubsc:NewModel}, there are two explanations for $\tau_R = (\tau_{\eta}\tau_0)^{1/2} \propto (\eta/T)^{1/2}$ and the interpretation of constant $\tau_0$ \cite{korb:jpcc2015,kausik:petro2019}, and more investigations are required to narrow down the theory.

The BPP model predicts a ``kink" in $T_2$ during the transition from the low- to high-viscosity regimes, where $T_2$ is shifted up by a factor 10/3 with increasing viscosity. The kink is supposed to occur at a viscosity corresponding to $\omega_0 \tau_{\eta} = 0.615$, which as shown in Fig. \ref{fg:T2crude} predicts am intermittent spread between low frequency (2.3 MHz) and high frequency (400 MHz) data. However, no spread between the 2.3 MHz and 400 MHz $T_{2LM}$ data is apparent (within uncertainties), for both polydisperse polymers and bitumen. In other words, there is no apparent frequency dependence in $T_{2LM}$ (within uncertainties) during the low- to high-viscosity regime, at least up to 400 MHz. We also note that such a kink in $T_{2LM}$ has never been reported before for polydisperse fluids with a broad $T_2$ distribution.

We also note that the MD simulations of the polymers agree well with the measurements, which again suggests that $^1$H-$^1$H dipole-dipole relaxation dominates over paramagnetism at high-viscosities.

\subsection{Full frequency dependence of polydisperse polymer}\label{subsc:FastMotion}

\begin{figure}[!ht]
	\begin{center} 
\includegraphics[width=1\columnwidth]{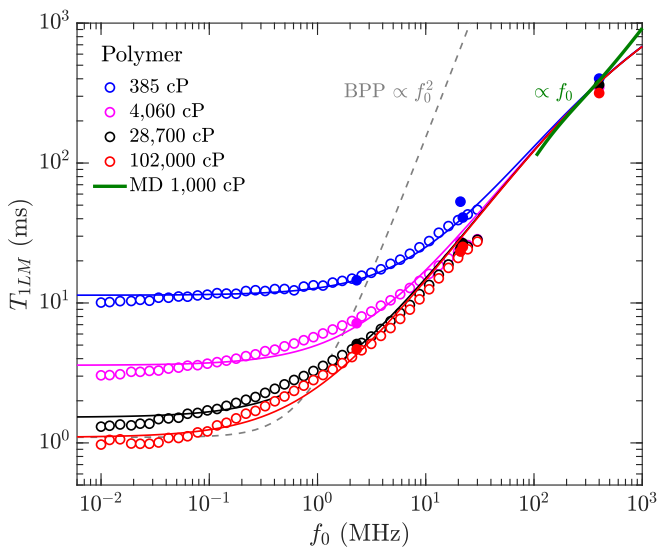} 
	\end{center}
	\caption{$T_{1LM}$ vs. $f_0$ from field-cycling (open symbols) and static fields (closed symbols) for the polydisperse polymers. Solid curves are results of the model in Eq. \ref{eq:NewLS} with fixed values of $S^2$ and $\tau_L$ listed in Table \ref{tb:SlowMotion}, along with optimized $\tau_0$ listed in Table \ref{tb:FastMotion}. MD simulations of the 1,000 cP polymer are shown above $f_0>$ 100 MHz, corresponding to viscosity independent region (see Fig. \ref{fg:MD_T1}) where $T_1\propto f_0$. BPP prediction where $T_1\propto f_0^2$ at high frequencies is also shown.}\label{fg:FC}
\end{figure}

\begin{figure}[!ht]
	\begin{center}
\includegraphics[width=0.94\columnwidth]{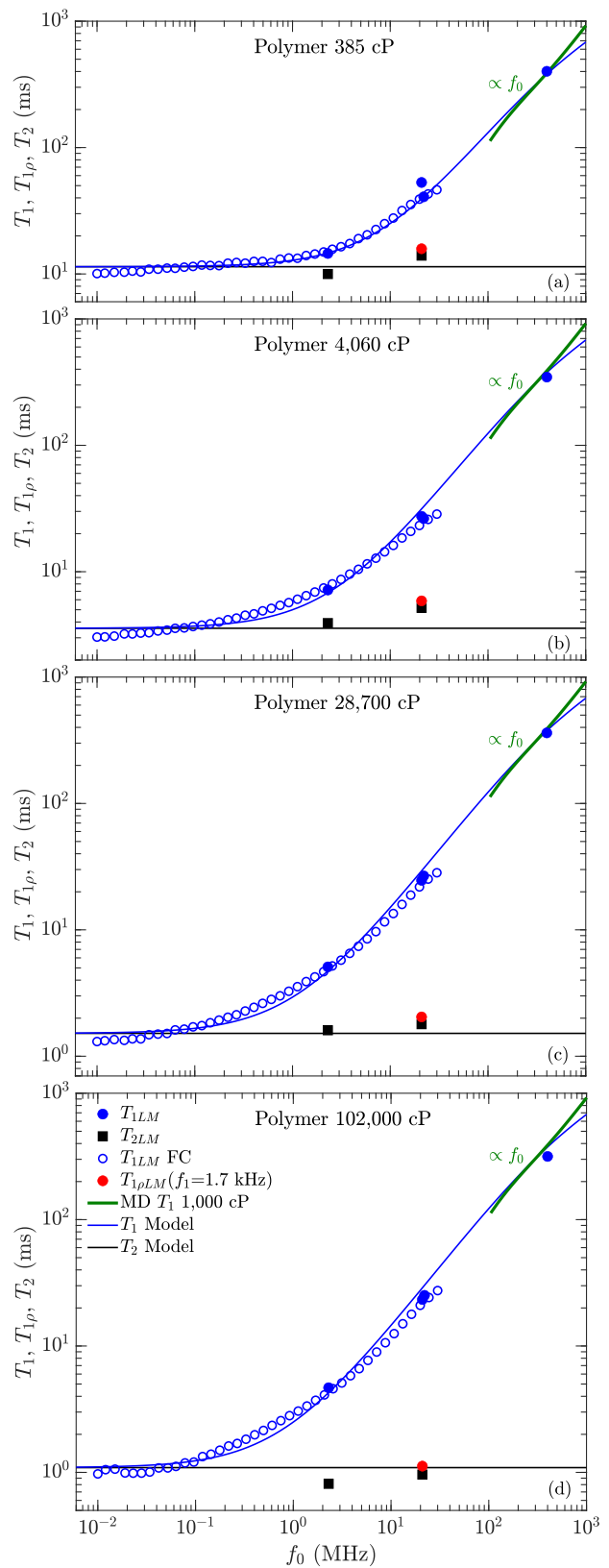} 
\caption{$T_{1LM}$ from Fig. \ref{fg:FC}, plus $T_2$, $T_{1\rho}$, and model.}\label{fg:OmegaPanel}
	\end{center}
\end{figure}

The full frequency dependence in $T_{1LM}$ for the polydisperse polymer are presented in Fig. \ref{fg:FC}. Also shown are the fits from the full expression of the model in Eq. \ref{eq:NewLS} using fixed values of $S^2$ and $\tau_L$ listed in Table \ref{tb:SlowMotion}, and optimized values of $\tau_0$ listed in Table \ref{tb:FastMotion}. The optimized values of $\tau_0$ in Table \ref{tb:FastMotion} tend to decrease with increasing viscosity, however the average agrees with the value in Table \ref{tb:SlowMotion} determined from the entire set of $T_{2LM}$ data. The factor $\simeq$ 2 smaller $\tau_0$ (and factor $\simeq \,2^{1/2}$ smaller $\tau_R$) for 102,000 cP is likely due to FC ramp-time effects which overestimate $T_{1LM}$ by a factor $\simeq \, 2^{1/2}$.

\begin{table}[!ht]
	\centering
	\begin{tabular}{ccc|cc}				
		\hline
		 $^{{\strut}{}}$ All at  & $\eta$ & $ \tau_{\eta}$ & $\tau_0$ & $\tau_R$  \\
		$^{{\strut}{}}$ 40$^{\circ}$C & (cP)&	 (ns) & (ns) & (ns) \\
		\hline
		$^{{\strut}{}}$& 385  & 2 $^{{\strut}{}}$&$^{{\strut}{}}$ 53& 11\\
		$^{{\strut}{}}$& 4,060 & 25 $^{{\strut}{}}$&$^{{\strut}{}}$ 51& 36 \\
		$^{{\strut}{}}$& 28,700 & 179 $^{{\strut}{}}$&$^{{\strut}{}}$ 40& 85 \\
		$^{{\strut}{}}$& 102,000 & 636 $^{{\strut}{}}$&$^{{\strut}{}}$ 22$^*$ &  118$^*$
	\end{tabular}
	\caption{Viscosity $\eta$, Stokes-Einstein-Debye correlation time $\tau_{\eta}$ (Eq. \ref{eq:BPP2}), $\tau_0$ fitting parameter used in Fig. \ref{fg:FC} and corresponding $\tau_R = \left(\tau_{\eta}\tau_0\right)^{1/2}$, for the polydisperse polymers at 40$^{\circ}$C. ($^*$) $\tau_0$ and $\tau_R$ for 102,000 cP are likely underestimated due to FC ramp-time effects.} \label{tb:FastMotion}
\end{table}

The full expression of the model shows good agreement with the data over the entire frequency range $f_0 = 0.01 \leftrightarrow 400$ MHz, including the coalescence at high frequencies corresponding to the slow-motion regime $\omega_0\tau_R \gg 1$. Likewise, the MD simulation of the 1,000 cP polymer above $f_0>$ 100 MHz shows $T_1\propto f_0$ behavior, consistent with the data and the model.

Fig. \ref{fg:OmegaPanel} shows the same $T_{1LM}$ data as Fig. \ref{fg:FC}, along with $T_{2LM}$ and $T_{1\rho LM}$ (at $f_1$ = 1.7 kHz) which show similar values to $T_{1LM}$ at $f_0$ = 0.01 MHz. This implies that $T_{2LM}$ and $T_{1\rho LM}$ ($f_1$ = 1.7 kHz) have no dependence on $f_0$ (within uncertainties), and remain in the fast-motion regime even when $\omega_0\tau_R \gg 1$. We speculate this is due to the broad distribution in correlation times, which for $T_1$ is more efficiently narrowed by longitudinal cross-relaxation (a.k.a. spin-diffusion) than for $T_2$ by transverse cross-relaxation. This stems from the fact that transverse cross-relaxation for $T_2$ is only effective in the ``non-secular" limit, i.e. the low frequency limit \cite{kowalewski:book}. As such, we speculate that the transition from the fast- to slow-motion regime may occur over a much broader frequency range for $T_2$ than for $T_1$. This is supported by the lack of frequency dependence (i.e. the lack of a kink) in $T_{2LM}$ shown in Fig. \ref{fg:T2crude}(b). As such, we retain the factor 10/3 in Eq. \ref{eq:LS_T2} even in the slow-motion regime, at least up to 400 MHz.

\section{Conclusions}\label{sc:Conclusions}

We present NMR relaxation measurements of polydisperse polymers over a wide range of viscosities and over a wide range of frequencies using $T_{1}$ and $T_2$ in static fields, $T_{1}$ field-cycling relaxometry, and $T_{1\rho}$ relaxation in the rotating frame.

We develop a phenomenological model to fit the relaxation data which accounts for a distribution in molecular correlation times for these polydisperse polymers by decreasing the frequency exponent in the BPP model \cite{bloembergen:pr1948} from $(\omega\tau)^2$ to $(\omega\tau)^1$ \cite{singer:SPWLA2017,singer:EF2018}. Our model also accounts for internal motions of the non-rigid polymer branches with a Lipari-Szabo model incorporating an order parameter $S^2$ (i.e. rigidity), and a (fast) local correlation time $\tau_L$ of the polymer branches. In the high viscosity regime of the model, the (slow) rotational correlation time of the entire polymer is taken to be $\tau_R = \left(\tau_{\eta}\tau_0\right)^{1/2} \propto (\eta/T)^{1/2}$, where $\tau_{\eta}\propto \eta/T$ is the Stokes-Einstein-Debye correlation time, and $\tau_0$ is a constant interpreted in \cite{korb:jpcc2015,kausik:petro2019}.

In the high-viscosity (i.e. slow-motion) regime, the model accounts for the viscosity independent ``plateau" $T_{1LM} \times 2.3/f_0 \simeq$ 3 ms of the polydisperse polymers. The model accounts for previously reported bitumen data where the same plateau was reported, as well as the departure from the $T_{1LM} \propto f_0$ behavior at high frequencies $f_0 \gtrsim$ 100 MHz. The model also accounts for the $T_{2LM} \propto (\eta/T)^{-1/2}$ behavior at high viscosities $\eta/T \gtrsim $ 0.3 cP/K (or $\eta \gtrsim $ 100 cP at ambient, equivalently), for both polydisperse polymers and bitumen. 

The model is applied to the full range of frequencies $f_0$ covering both the fast- and slow-motion regimes of the polydisperse polymers. The data indicate that the $T_{2LM}$ and $T_{1\rho}$ are independent of $f_0$ up to 400 MHz, which we speculate is because the transition from the fast- to slow-motion regime occurs over a much broader frequency range for $T_2$ than for $T_1$. We speculate this may be a result of the broad distribution in correlation times together with cross-relaxation (a.k.a. spin-diffusion) effects. 

Molecular dynamics simulations of $T_1$ and $T_2$ by $^1$H-$^1$H dipole-dipole relaxation of monodisperse polymers are reported with viscosities in the range $\eta = 1 \leftrightarrow $ 1,000 cP. The simulations confirm the dominance of int{\it ra}-molecular over int{\it er}-molecular $^1$H-$^1$H relaxation at high viscosity, which was previously only assumed to be the case. The simulations for $\eta \gtrsim$ 100 cP show that $T_{1} \propto f_0$ at high frequencies ($f_0 \gtrsim 100 $ MHz), specifically $T_{1} \times 2.3/f_0 \simeq$ 3 ms, in good agreement with measurements and the model. A similar dispersion relation $T_{1S} \propto f_0$ was previously reported from MD simulations of the surface relaxation $T_{1S}$ of heptane confined in a polymer matrix, specifically $T_{1S} \times 2.3/f_0 \simeq$ 7 ms \cite{parambathu:arxiv2020}, implying a common NMR relaxation mechanism between high-viscosity fluids and low-viscosity fluids under confinement. The MD simulations also imply that $^1$H-$^1$H dipole-dipole relaxation dominates over paramagnetism for these systems.

\section*{Supporting information} 
\begin{enumerate} [noitemsep]
	\item Effects of dissolved oxygen.
	\item Details of the new model.
	\item Link between model and Cole-Davidson.
	\item Details of the MD simulation.
\end{enumerate} 
\section*{Acknowledgments} 

We thank the Chevron Corporation, the Rice University Consortium on Processes in Porous Media, and the American Chemical Society Petroleum Research Fund (No. ACS PRF 58859-ND6) for funding this work. We gratefully acknowledge the National Energy Research Scientific Computing Center, which is supported by the Office of Science of the U.S. Department of Energy (No.\ DE-AC02-05CH11231) and the Texas Advanced Computing Center (TACC) at The University of Texas at Austin for HPC time and support. We also thank Zeliang Chen, Maura Puerto, Lawrence B. Alemany, Kairan Zhu, Z. Harry Xie, Tuan D. Vo, Prof. Aydin Babakhani, Prof. Rafael Verduzco, Hao Mei, and Jinlu Liu for their contributions to Ref. \cite{singer:EF2018}, which paved the way for the further investigations reported here.

\clearpage

\section{Supporting Information}

\subsection{Effects of dissolved oxygen}\label{sup:Oxygen}

The measured $T_{1,2}$ of a bulk fluid at atmospheric conditions are given by the following:
\begin{align}
\frac{1}{T_{1,2}} &= \frac{1}{T_{1B,2B}} + \frac{1}{T_{1{\rm O_2},2{\rm O_2}}}
\label{eq:O2}
\end{align}
$T_{1B,2B}$ are the bulk contributions from $^1$H-$^1$H dipole-dipole relaxation. For de-oxygenated $n$-alkanes at ambient it was previously shown that $T_{1B,2B} = 9.56\,(\eta/T)^{-1}$ \cite{lo:SPE2002}, which for $n$-heptane is $T_{1B,2B}$ = 7,320 ms at 25 $^{\circ}$C (where $\eta$ = 0.39 cP).

\begin{figure}[!ht]
	\begin{center}
\includegraphics[width=1\columnwidth]{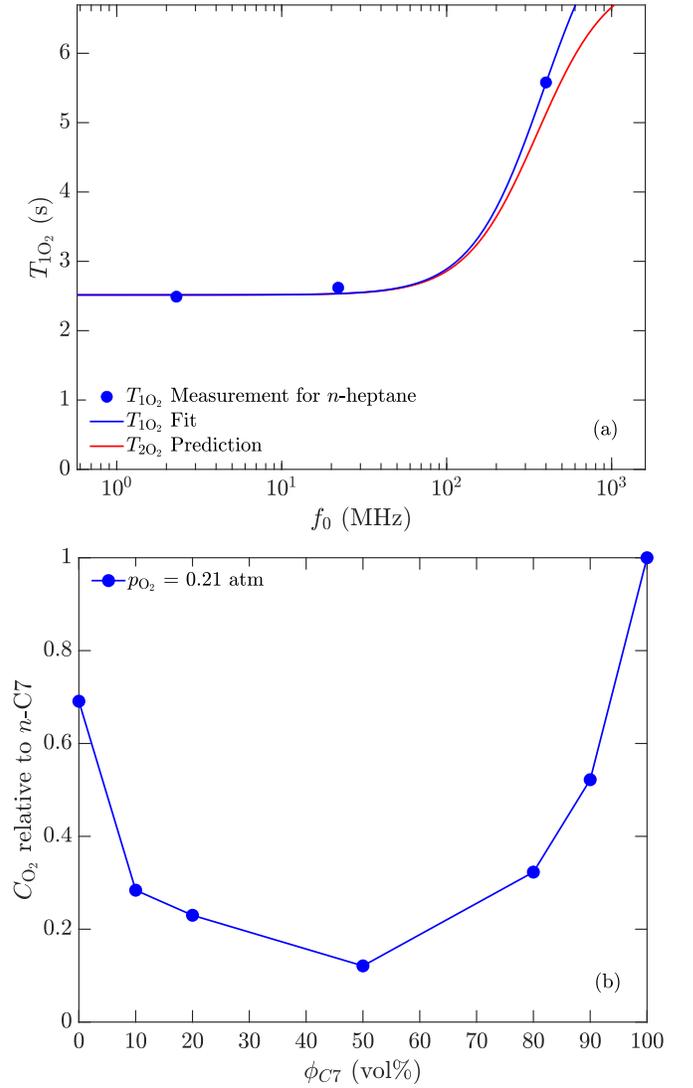} 
\end{center}
	\caption{(a) Measurement of paramagnetic relaxation time $T_{1{\rm O_2}}$ due to dissolved oxygen in $n$-heptane \cite{singer:EF2018} determined from Eq. \ref{eq:O2}, along with fit using Eq. \ref{eq:JO2}. Prediction for $T_{2{\rm O_2}}$ also shown. (b) MD simulation of concentration $C_{\rm O_2}$ of dissolved oxygen in 1,000 cP polymer ($\phi_{C7} = 0$ vol\%) relative to $n$-heptane ($\phi_{C7} = 100$ vol\%) under ambient conditions, along with simulations for various polymer-heptane mixes, taken from \cite{parambathu:arxiv2020}.} \label{fg:Oxygen}
\end{figure}

Given the measured $T_{1,2}$ and the known $T_{1B,2B}$ for $n$-heptane at ambient, Eq. \ref{eq:O2} yields the paramagnetic contribution $T_{1{\rm O_2}}$ from dissolved oxygen. As shown in Fig. \ref{fg:Oxygen}(a), $T_{1{\rm O_2}}$ increases from $T_{1{\rm O_2}} = $ 2,500 ms at $f_0 = 2.3 $ MHz to $T_{1{\rm O_2}} =  $ 5,600 ms at $f_0 = 400 $ MHz. $T_{1{\rm O_2}}$ depends on the electron correlation time $\tau_e$ and the concentration $C_{\rm O_2}$ of dissolved oxygen through the following relations \cite{mcconnell:book}:
\begin{align}
\frac{1}{T_{1{\rm O_2}}} &= J_{{\rm O_2}}(\omega_0)  + \frac{7}{3} J_{{\rm O_2}}(\omega_e), \label{eq:JO2}\\
\frac{1}{T_{2{\rm O_2}}} &= \frac{2}{3}J_{{\rm O_2}}(0)  + \frac{1}{2}J_{{\rm O_2}}(\omega_0)  + \frac{13}{6} J_{{\rm O_2}}(\omega_e), 
\end{align}
where the BPP spectral density is given by:
\begin{align}
J_{{\rm O_2}}(\omega)  &= \frac{1}{3} \Delta\omega^2_{\rm O_2} \frac{2 \tau_e}{1+(\omega \tau_e)^2} \,\,\, {\rm with} \,\,\, \Delta\omega^2_{\rm O_2} \propto C_{\rm O_2}\label{eq:JO2b} 
\end{align}
$\tau_e$ is the electron correlation time of the paramagnetic O$_2$ molecule, and the second moment (i.e. strength) is given by $\Delta\omega^2_{\rm O_2} \propto C_{\rm O_2}$. Note that the Larmor frequency of the electron $\omega_e$ is larger than $^1$H by a factor $\omega_e \simeq 659\, \omega_0$. The BPP model for the spectral density $J_{{\rm O_2}}(\omega)$ was previously shown to work well for a variety of solvents of various molecular weights \cite{teng:jmr2001}. Fig. \ref{fg:Oxygen}(a) shows that the fit to $T_{1{\rm O_2}}$ using Eqs. \ref{eq:JO2} and \ref{eq:JO2b} work well, with best fit parameters $\tau_e = 1.15$ ps (which is consistent with \cite{teng:jmr2001}) and $\Delta\omega_{\rm O_2}/2\pi $ = 62.8 kHz.

Given that the concentration of dissolved oxygen in $n$-heptane is  $C_{\rm O_2} = $ 132 mg/L (or 4.12 mM equivalently) at ambient ($p_{\rm O_2}$ = 0.21 atm) implies that the relaxivity (defined in MRI terminology \cite{lauffer:cr1987}) of O$_2$ is $r_1 = 0.10 $ mM$^{-1}$s$^{-1}$ at low frequencies $f_0\lesssim $ 100 MHz.

As shown in \cite{teng:jmr2001}, $\tau_e$ is roughly constant for solvents with the molecular weight of heptane or higher, and also appears roughly constant at lower molecular weights \cite{ward:jpc2018}. Furthermore as shown in Fig. \ref{fg:Oxygen}(b), MD simulations indicate that $C_{\rm O_2}$ is $\sim$30 \% less for the 1,000 cP polymer than for heptane \cite{parambathu:arxiv2020}, implying that $T_{1{\rm O_2}}$ is $\sim$30 \% larger for the polymer than for heptane. This predicts that $T_{1{\rm O_2}} \simeq  $ 8,000 ms for the polymer at $f_0 = 400 $ MHz, which is much larger than the measured $T_{1} \simeq  $ 300 ms. In other words, the effects of oxygen on $T_{1,2}$ are negligible for the polymer at 400 MHz. Below 400 MHz, the effects of oxygen on the polymer are even less since the measured $T_{1,2} \ll  $ 300 ms.

The only cases in this report where the effects of dissolved oxygen are apparent are for the low-viscosity crude-oils at $f_0 \lesssim $ 100 MHz where $T_{1LM,2LM} \gtrsim $ 1,000 ms. In such cases $T_{1LM,2LM}$ deviate slightly from the BPP correlation line $T_{1LM,2LM} = 4.0 \left(\eta/T\right)^{-1}$ in the low-viscosity regime.

\clearpage

\subsection{Details of the model}\label{sup:Model}

\begin{figure}[!ht]
	\begin{center}
\includegraphics[width=1\columnwidth]{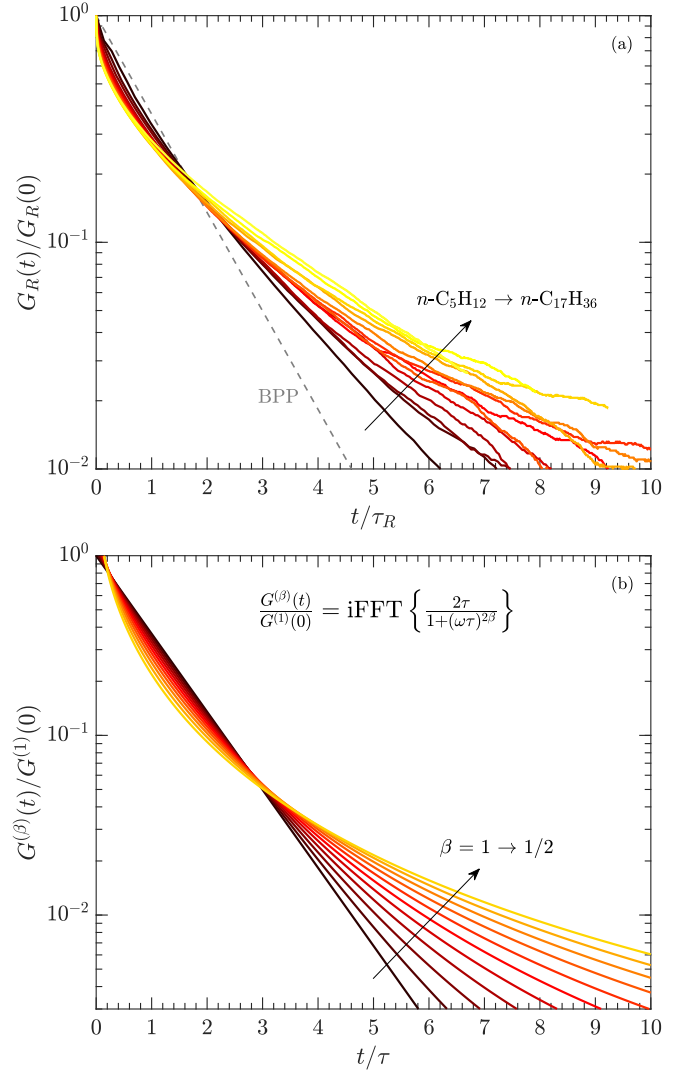} 	
	\end{center}
	\caption{(a) MD simulations of the intra-molecular auto-correlation function $G_R(t)$ for the liquid $n$-alkanes at ambient, taken from \cite{singer:jmr2017}. The $y$-axis is normalized by the zero time amplitude $G_R(0)$, and the $x$-axis is normalized by the correlation time $\tau_R$. Straight line is the BPP prediction. (b) Inverse Fourier transform $G^{(\beta)}(t)$ (Eq. \ref{eq:Beta2}) of the spectral density function $J^{(\beta)}(\omega)$ (Eq. \ref{eq:Beta1}), where $\beta = 1$ corresponds to BPP model (Eq. \ref{eq:Gt1}), and $\beta = 1/2$ (Eq. \ref{eq:Gt2}) corresponds to the model in this report.} \label{fg:GTcomparison}
\end{figure}

The autocorrelation function $G(t)$ for fluctuating magnetic $^1$H-$^1$H dipole-dipole interactions is central to the development of the NMR relaxation theory in liquids \cite{bloembergen:pr1948,torrey:pr1953,abragam:book,hwang:JCP1975,mcconnell:book,cowan:book,kimmich:book}. From $G(t)$, one can determine the spectral density function $J(\omega)$ by Fourier transform as such:
\begin{equation}
J(\omega) = 2\int_{0}^{\infty}G(t)\cos\left(\omega t\right) dt,
\label{eq:FourierRTcos}
\end{equation}
for $G(t)$ in units of s$^{-2}$ \cite{mcconnell:book}. As shown in Fig. \ref{fg:GTcomparison}(a), $G(t)$ for the $n$-alkanes has a multi-exponential (i.e. ``stretched") decay rather than the single exponential decay predicted by BPP. Furthermore, the degree of stretching increases with increasing carbon number (i.e. increasingly non-spherical geometry). MD simulations indeed confirm that $n$-pentane and $n$-hexane have a more stretched intra-molecular $G_R(t)$ autocorrelation function (i.e. a broader distribution of correlation times) than the symmetric neo-pentane and benzene molecules, respectively \cite{singer:jcp2018}. The degree of stretching can be further analyzed using inverse Laplace transforms (see also \cite{singer:jcp2018b} and supporting information in \cite{singer:jcp2018}).

By comparison, Fig \ref{fg:GTcomparison}(b) shows the effect of decreasing the exponent $\beta$ in the spectral density function $J^{\left(\beta\right)}\!(\omega)$ as such:
\begin{align}
J^{\left(\beta\right)}\!(\omega) &= \frac{1}{3} \Delta\omega_R^2  \frac{2 \tau}{1+\left(\omega\tau\right)^{2\beta}} \label{eq:Beta1},\\
G^{\left(\beta\right)}\!(t) &=  \frac{2}{2\pi}\int_{0}^{\infty}J^{\left(\beta\right)}\!(\omega) \cos\left(\omega t\right) d\omega,
\label{eq:Beta2}
\end{align}
where the real part of the inverse Fourier transform $G^{\left(\beta\right)}\!(t)$ is also defined. An analytical expression for $G^{(\beta)}(t)$ exists for the two extreme cases:
\begin{align}
\frac{G^{\left(1\right)}(t)}{G^{\left(1\right)}(0)} &=  \exp\left(-t/\tau\right), \label{eq:Gt1}\\
\frac{G^{\left(1/2\right)}(t)}{G^{\left(1\right)}(0)}  &=\frac{2}{\pi}\int_{0}^{\infty}\frac{\cos(x)}{x + t/\tau}dx. \label{eq:Gt2}
\end{align}
The case of $\beta = 1$, $G^{\left(1\right)}\!(t)$, is the BPP model with a single-exponential decay. The case of $\beta = 1/2$, $G^{\left(1/2\right)}\!(t)$ \cite{abramowitz:1964}, is the stretched case used in the phenomenological model to account for the independence of $T_{1LM}$ on $\tau$, i.e. the independence on $\eta/T$. As demonstrated in Fig \ref{fg:GTcomparison}, increasing the carbon number increases the degree of stretching in Fig \ref{fg:GTcomparison}(a), which corresponds to decreasing $\beta$ in Fig \ref{fg:GTcomparison}(b). 

It is interesting to note that the long time behavior of Eq. \ref{eq:Gt2} is $G^{\left(1/2\right)}\!(t) \propto t^{-2}$ for $t/\tau \gg 10$. This is analogous to the Mittag-Leffler function, which starts off as a stretched exponential at short times, and turns into a power-law decay at long times \cite{magin:jmr2011}.

We note that $G^{\left(1/2\right)}\!(t)$ starts to diverge below $t/\tau = 0.14$, therefore the model is valid for times $t/\tau \gtrsim 0.14$. Correspondingly, this implies that the model for $J^{\left(1/2\right)}\!(\omega)$ is valid for frequencies $2 \omega \tau \lesssim 7.1$ (where the factor 2 reflects the $J(2 \omega_0)$ terms in Eq. \ref{eq:T12J}). Given the local correlation-time $\tau_L \simeq 50$ ps, this corresponds to validity below frequencies $f_0 \lesssim $ 10 GHz.

\subsection{Link between model and Cole-Davidson}\label{supB:Cole}

The Cole-Davidson function is widely used to describe the dielectric response \cite{davidson:jcp1951,lindesy:jcp1980,beckmann:prep1988} and its departure from the traditional Debye model. The function is defined as such:
\begin{align}
\chi''(\omega) &=  {\rm Im} \left\{\left(\frac{1}{1+i\omega\tau_{CD}}\right)^{\!\beta_{CD}}\!\right\} \label{eq:Chi1}
\end{align}
which reduces to the following expression:
\begin{align}
\chi''(\omega)  &=  \frac{\sin\left[\beta_{CD}\arctan(\omega\tau_{CD})\right]}{\left[1+(\omega\tau_{CD})^2\right]^{\beta_{CD}/2}}.
\label{eq:Chi2}
\end{align}
The Cole-Davidson exponent is bound by $0 < \beta_{CD} \leq1$, where $\beta_{CD} = 1$ is the Debye model. As $\beta_{CD}$ is decreased, the underlying distribution $P_{CD}(\tau)$ in molecular correlation times $\tau$ grows wider. 

\begin{figure}[!ht]
	\begin{center}
\includegraphics[width=1\columnwidth]{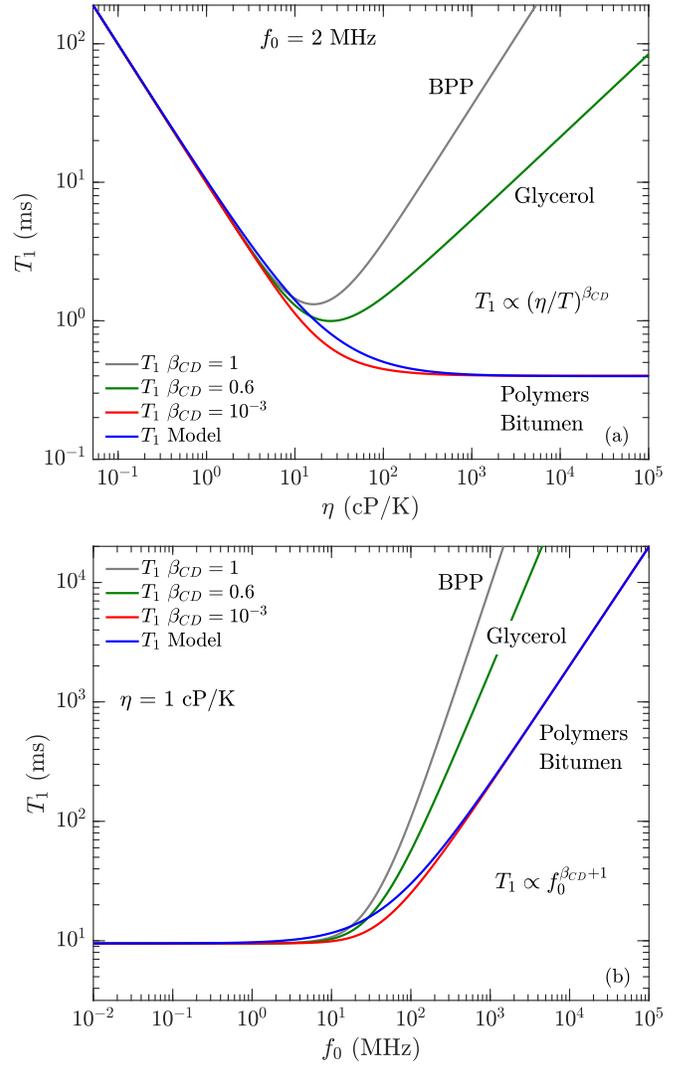}
\end{center}
	\caption{$T_1$ determined from $J_{CD}(\omega)$ for $\beta_{CD}$ = 1, 0.6, and $10^{-3}$, compared with our model, (a) as a function of $\eta/T$ for fixed $f_0$ = 2 MHz, and (b) as a function of $f_0$ for fixed $\eta/T$ = 1 cP/K.} \label{fg:CD}
\end{figure}

The Cole-Davidson function has also been successfully used to describe NMR relaxation of glycerol and monodispersed polymers through the relation $\chi_{DD}''(\omega) \propto \omega/T_1 $ \cite{kruk:pnmrs2012,flamig:jpcb2020}. The NMR Cole-Davidson spectral density $J_{CD}(\omega)$ is given by the following:
\begin{align}
J_{CD}(\omega) &= \frac{2}{\omega} {\rm Im} \left\{\left(\frac{1}{1+i\omega\tau_{CD}}\!\right)^{\!\beta_{CD}}\right\}\frac{\frac{1}{3}\Delta\omega_R^2}{\beta_{CD}\pi/2} \label{eq:Chi3}
\end{align}
which reduces to the following expression:
\begin{align}
J_{CD}(\omega) &= \frac{2}{\omega} \frac{\sin\left[\beta_{CD}\arctan(\omega\tau_{CD})\right]}{\left[1+(\omega\tau_{CD})^2\right]^{\beta_{CD}/2}} \frac{\frac{1}{3}\Delta\omega_R^2}{\beta_{CD}\pi/2}.  \label{eq:Chi4}
\end{align}
For the purposes of comparison with our model, we have added a factor $\frac{1}{3}\Delta\omega_R^2/\beta_{CD}\pi/2$ to the definition of $J_{CD}(\omega)$. The factor $\frac{1}{3}\Delta\omega_R^2$ is related to the units convention of the spectral density \cite{cowan:book}. The factor $1/\beta_{CD}\pi/2$ is introduced to highlight the similarity with our model. $J_{CD}(\omega)$ as defined in Eq. \ref{eq:Chi4} becomes independent of $\beta_{CD}$ in the limit $\beta_{CD} \lesssim 10^{-3}$, and tends towards:
\begin{align}
J_{CD}(\omega) & \Rightarrow \frac{1}{3}\Delta\omega_R^2 \frac{2 \tau_R}{1+\omega\tau_R}\,\,\,\,{\rm for} \,\,
\begin{cases}
    \beta_{CD} \, \lesssim \, 10^{-3}\\
   \tau_{CD} = \tau_R \,\pi/2             
\end{cases} \label{eq:Cole}
\end{align}
In other words, $J_{CD}(\omega)$ tends towards our model (without the added Lipari-Szabo model, i.e. $S^2 = 1$) in the case of $\beta_{CD}\lesssim 10^{-3}$ and $\tau_{CD} = \tau_R\,\pi/2 $. Fig \ref{fg:CD} shows $T_1$ determined from $J_{CD}(\omega)$ for $\beta_{CD}$ = 1, 0.6, and $10^{-3}$, compared with our model. The similarity between $J_{CD}(\omega)$ for $\beta_{CD} \lesssim 10^{-3}$ and our model is remarkable, and only shows a $\simeq$ 22\% deviation at the transition from the fast- to slow-motion regime. The deviation for $T_2$ (not shown) $\simeq$ 14\% is even less.

Fig \ref{fg:CD} also indicates the corresponding cases as BPP ($\beta_{CD}$ = 1), glycerol ($\beta_{CD} \simeq$ 0.6 for the $\alpha$-peak  \cite{davidson:jcp1951,kruk:pnmrs2012,flamig:jpcb2020}), and our model for polydisperse polymers and bitumen ($\beta_{CD} \lesssim 10^{-3}$). Also shown are the limiting behavior in according to Cole-Davidson in the slow-motion regime $T_1 \propto f_0^{\beta_{CD}+1} \left(\eta/T\right)^{\beta_{CD}}$.

The above comparison shows that polydisperse polymers and bitumen have a much larger distribution in molecular correlation times than gycerol. In terms of the Cole-Davidson model, this manifests itself as a small exponent $\beta_{CD}\lesssim 10^{-3}$ for the polydisperse polymers and bitumen, compared to monodisperse gycerol where $\beta_{CD}\simeq 0.6$ for the $\alpha$-peak.

\clearpage

\subsection{Details of MD simulation}\label{sup:MD}

The detailed simulation procedure is provided in Ref. \cite{parambathu:arxiv2020}. The system was simulated using NAMD \cite{phillips:JCC2005,NAMD} code, and modeled using CGenFF Force Field\cite{cgenff:2010,cgenff:2020}. The number of molecules used were 269, 143, 74, and 40 for the 2-mer, 4-mer, 8-mer and 16-mer, respectively. The systems were equilibrated at constant $NpT$ conditions at 25$^{\circ}$C and 1 atm. The data was then collected in a $NVE$ ensemble to compute the autocorrelations $G_{R,T}(t)$.

\begin{figure}[!ht]
	\begin{center}
\includegraphics[width=1\columnwidth]{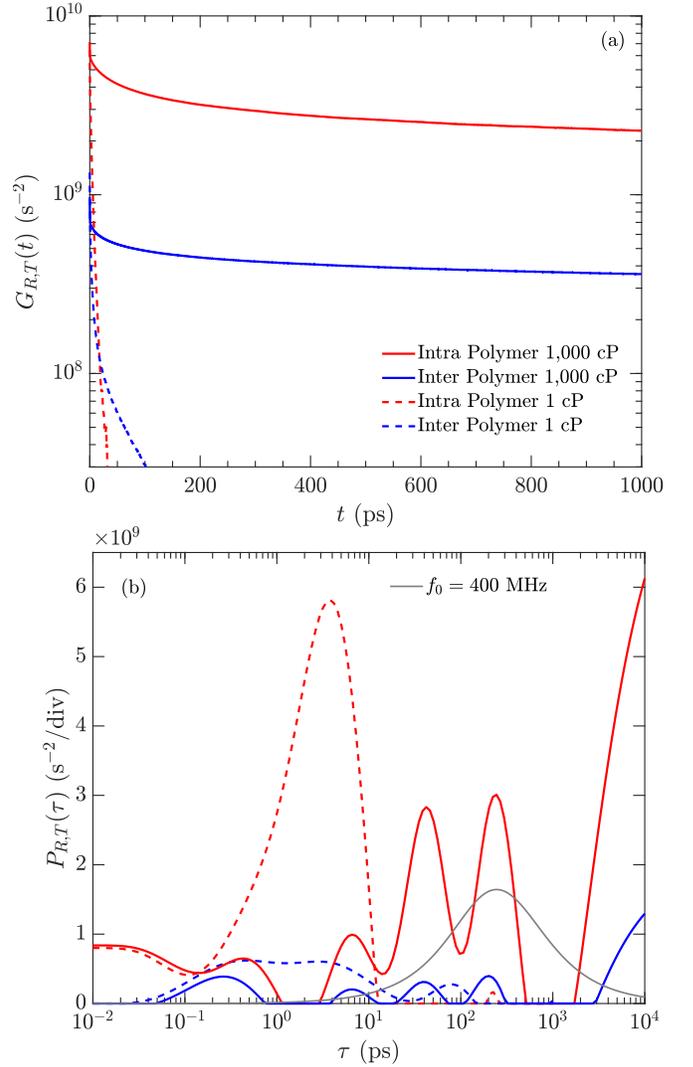}
\end{center}
	\caption{(a) MD simulations of the intra-molecular ($G_R(t)$) and inter-molecular ($G_T(t)$) auto-correlation functions $G_{R,T}(t)$ for polymer at 1,000 cP and 1 cP. (b) Probability distributions $P_{R,T}(\tau)$ determined from inverse Laplace transforms of $G_{R,T}(t)$ (Eq. \ref{eq:ILT1}). Gray curve is BPP frequency filter defined in Eq. \ref{eq:JwfromPt}, at $f_0 = $ 400 MHz.} \label{fg:ILT}
\end{figure}

The results of the intra-molecular $G_R(t)$ and inter-molecular $G_T(t)$ are shown in Fig. \ref{fg:ILT}(a) for the polymer at 1 cP and 1,000 cP. In order to quantify the departure of $G_{R,T}(t)$ from single-exponential decay, we fit $G_{R,T}(t)$ to a sum of multi-exponential decays and determine the underlying probability distribution $P_{R,T}(\tau)$ in correlation times $\tau$. More specifically, we perform an inversion of the following Laplace transform \cite{venkataramanan:ieee2002,song:jmr2002,singer:jcp2018}: 
\begin{align}
G_{R,T}(t) &= \int_{0}^{\infty}\! P_{R,T}(\tau) \exp\left(-t/\tau\right) d\tau, \label{eq:ILT1}\\
\tau_{R,T} &= \frac{1}{G_{R,T}(0)}\int_{0}^{\infty} \! P_{R,T}(\tau)\, \tau \, d\tau,\label{eq:ILT2} \\
G_{R,T}(0) &= \frac{1}{3} \Delta\omega_{R,T}^2
\end{align}
where $P_{R,T}(\tau)$ are the probability distribution functions derived from the inversion, plotted in Fig. \ref{fg:ILT}(b). Details of the inversion procedure can be found in \cite{singer:jcp2018b} and in the supporting information in \cite{singer:jcp2018}.

The $P_{R,T}(\tau)$ in Fig. \ref{fg:ILT}(b) indicate a set of $\sim$5 polymer modes, located at similar $\tau$ values for both intra-molecular $P_{R}(\tau)$ and inter-molecular $P_{T}(\tau)$ interactions. The intra-molecular $P_{R}(\tau)$ has an additional mode at short $\tau\sim 10^{-2}$ ps for both the polymer and heptane, while it is absent for $P_{T}(\tau)$ in both cases. Similar observation of the $\tau\sim 10^{-2}$ ps mode was reported in the supporting information for all the liquid-state $n$-alkanes \cite{singer:jcp2018}, and it is attributed to the fast rotation of the methyl groups.

The decomposition of $G_{R,T}(t)$ into a sum of exponential decays is common practice in phenomenological models of complex molecules \cite{beckmann:prep1988,bakhmutov:book}, where the more complex the molecular dynamics, the more exponential terms are required \cite{woessner:jcp1962,woessner:jcp1965}. This is in contrast to analytical techniques and theories used to interpret the autocorrelation function of monodisperse polymers with high molecular-weight $M_w > $ 4,000 g/mol where entanglement occurs, and a power law decay is observed $G(t) \propto t^{-\alpha}$ \cite{chavez:macro2011,chavez:macro2011b,mordvinkin:jcp2017}. We note however that our polydisperse polymers most likely do not entangle to the extent that monodisperse polymers do, if indeed the polydisperse polymers entangle at all.

Also defined in Eq. \ref{eq:ILT2} are the correlation times $\tau_{R,T}$ derived from $P_{R,T}(\tau)$, which are found to be $\tau_{R}$ = 2,380 ps and $\tau_{T}$ = 3,030 ps at 1,000 cP, compared with $\tau_{R}$ = 3.33 ps and $\tau_{T}$ = 10.1 ps at 1 cP. In other words, the MD simulations indicate that $\tau_{R,T}$ increase by a factor 
of $\simeq$500 from 1 cP to 1,000 cP. 

The spectral density $J_{R,T}(\omega)$ is determined from the Fourier transform (Eq. \ref{eq:FourierRTcos}) of $G_{R,T}(t)$ (Eq. \ref{eq:ILT1}):
\begin{align}
J_{R,T}(\omega) &= \int_{0}^{\infty}\! \frac{2\tau}{1+(\omega\tau)^2}   P_{R,T}(\tau) d\tau \label{eq:JwfromPt}
\end{align}
from which the $T_{1,2}$ dispersion (i.e. frequency dependence) can be determined as such:
\begin{align}
\frac{1}{T_{1,R,T}} &= J_{R,T}(\omega_0) + 4 J_{R,T}(2\omega_0),  \\
\frac{1}{T_{2,R,T}} &= \frac{3}{2} J_{R,T}(0) + \frac{5}{2} J_{R,T}(\omega_0) + J_{R,T}(2\omega_0),  \\
\frac{1}{T_{1,2}} &= \frac{1}{T_{1R,2R}} + \frac{1}{T_{1T,2T}}. \label{eq:T12}
\end{align}

Fig. \ref{fg:ILT}(a) indicates that the second moment $ \Delta\omega_{R,T}^2 = 3\,G_{R,T}(0)$ is about a factor $\simeq$10 larger for intra-molecular ($ \Delta\omega_{R}^2$) versus inter-molecular ($ \Delta\omega_{R}^2$) interactions at 1,000 cP. Given that $\tau_R \simeq \tau_T$ at 1,000 cP, one can then deduce that the ratio $T_{1T,2T}/T_{1R,2R} \simeq 10$. In other words, the intra-molecular relaxation rate is $\simeq 10$ larger than the inter-molecular relaxation rate, therefore intra-molecular relaxation dominates at 1,000 cP.

The gray curve in Fig. \ref{fg:ILT}(b) corresponds to the ``BPP frequency filter" defined in Eq. \ref{eq:JwfromPt}, at $f_0 = $ 400 MHz (as an example). In other words, the components of $P_{R,T}(\tau)$ contributing to $T_1$ at $f_0 = $ 400 MHz are weighted by the BPP frequency filter curve, which peaks at $\omega_0\tau = 0.615$. As such, the components in $P_{R,T}(\tau)$ at long $\tau \simeq 10^4$ ps do not contribute to $T_1$ at $f_0 = $ 400 MHz (as an example), nor do the short components $\tau \lesssim 10$ ps.

\clearpage

\end{document}